\documentclass[11pt,a4paper]{article}
\usepackage{epsfig}

\newlength{\dinwidth}
\newlength{\dinmargin}
\def\eq#1{{Eq.~(\ref{#1})}}
\newcommand{\Le}{\left(}
\newcommand{\Ra}{\right)}

\newcommand{\beq}{\begin{equation}}
\newcommand{\eeq}{\end{equation}}
\newcommand{\beqar}{\begin{eqnarray}}
\newcommand{\eeqar}{\end{eqnarray}}
\newcommand{\D}{\partial}

\newcommand{\as}{\alpha_s}

\newcommand{\Ph}{\Phi}
\newcommand{\Php}{\Phi^{\dagger}}
\newcommand{\php}{\phi^{\dagger}}

\newcommand{\rad}{r^{2}_{12}}
\newcommand{\radd}{r^{2}_{23}}
\newcommand{\raddd}{r^{2}_{31}}

\begin{document}

\title {{~}\\
{\Large \bf Langevin equation in effective theory \\
of interacting QCD pomerons in the limit of large $N_c$}\\}
\,
\author{ {~}\\
{~}\\
{\bf S. Bondarenko\,
\thanks{Email: sergey@fpaxp1.usc.es}}
 \\[10mm]
 {\it\normalsize University of Santiago de Compostela,}\\
 {\it\normalsize Santiago de Compostela, Spain}\\}
\maketitle
\thispagestyle{empty}

\begin{abstract}
 Effective field theory of interacting BFKL pomerons
is investigated and Langevin equation for the theory, which 
arises  
after the introduction of additional auxiliary field,  
is obtained. The Langevin
equations are considered for the case
of interacting BFKL pomerons
with  
both splitting and merging vertexes and for the 
interaction which includes
additional "toy" four pomeron interaction vertex. In the latest case
an analogy with the Regge field theory 
in zero dimensions (RFT-0) was used in order to 
obtain this  "toy" vertex, which coincided
with the four point function of two-dimensional
conformal field theory obtained in \cite{Korch1}.
The comparison between the Langevin equations
obtained in the frameworks of dipole and RFT approaches
is performed, the
interpretation of  results is given and possible 
application of obtained equations is discussed.

\end{abstract}

\section{Introduction}

 The scattering process of  hadrons and nuclei in QCD with a large number of
colors $N_c$ have been vigorously investigated in recent years 
along two main lines. 
The Color Glass Condensate (CGC) approach \cite{balitsky,jimwalk}, 
is formulated in the transverse position space and in large $N_c$ 
limit the evolution dynamics of the model is analyzed in the terms
of color dipoles  \cite{dipmod}. Another approach, 
the QCD Reggeon Field Theory (QCD-RFT), was written 
and investigated in the momentum space for 
the BFKL pomerons, \cite{bfkl1,bfkl2,bfkl3},
considering 
the picture of interaction of the pomerons in t-channel.
The history of developing of such a picture 
begins from the pioneric paper \cite{Gribov} and it bases 
on the standard diagrammatic calculus developed
for the interacting BFKL pomerons,
having , as well, transverse position space 
formulation in large $N_c$ limit of the theory, see  
\cite{vert1,eglla2,braun1,braun2,braun3,braun4}. 
Both approaches are expected to be 
valid also in  the case of symmetrical treatment of target
and projectile and may include not only a vertex of pomeron splitting but
also a vertex of merging of two pomerons in one. 
It gives the possibility for attempts to formulate and calculate
the pomeron loops contribution to the scattering amplitude
and makes these approaches formally similar.
But, in spite of this
formal similarity,  these approaches are written and formulated 
in the different frameworks and , therefore, the 
identity of the approaches is not fully clear.

 The QCD-RFT formulation of the high energy scattering
uses a Lagrangian language describing the processes of scattering
at high energy, \cite{braun1,braun2,braun3,braun4}, 
treating the 
processes of the high energy scattering in the terms
of nonlocal effective field theory , which is based on the conformal invariant
propagator of the BFKL pomerons and identical vertexes of 
pomeron splitting and merging. This approach has a long  history 
of developing in the form of fenomenological RFT, 
\cite{0dim1,0dim2}, 
and keeps a lot of mutual features with this "older brother",
see \cite{bond1}. The  dipole CGC and so-called JIMLWK approaches
are based on the consideration of the color dipoles as a main degrees 
of freedom in the 
high energy scattering in limit of large $N_c$ 
, \cite{jimwalk,dipmod}. It is proven , that on the level of the
zero transverse dimensions both approaches describes the same physics, 
\cite{bond2}, in spite of the fact that 
these two approaches use very different
pictures for the description of the scattering process. 
In QCD-RFT the calculations are based on the picture of 
t-channel propagating 
and interacting pomerons, whereas the  CGC approach uses the picture
of evolution of the dipoles 
of target and projectile in s-channel
with interactions of dipole showers after the evolution.
Therefore, 
considering the problem of  
pomeron's loops contribution into the scattering amplitude, 
both theories support
very different strategies for the accounting of the loops.
In  QCD-RFT, because of the Lagrangian formulation of the theory, 
a program of such calculation may be formulated
as usual  perturbative calculations and these calculations 
in general must
be very difficult, as it is always happens for the loops calculations
in a field theory. 
In CGC approach the problem of the calculations
of the pomeron's loops usually formulated as a problem of evolution
equation with both , merging and splitting vertexes included
and may be explicitly formulated in the terms of some effective 
Hamiltonian \cite{selfdual}.
A Hamiltonian formulation of the theory 
has been considered in QCD-RFT as well,
\cite{braun4},
but the effective Hamiltonian formulation of the high energy scattering processes
is out of the scope of the paper.

 In this paper we will investigate
another possible way to account the  pomeron loops in the
different high energy scattering approaches, namely a approach when 
loops contribution are  compactly coded in the Langevin equation.
On the level of CGC approach the different types 
of arising Langevin equations were considered in
\cite{stoch1,stoch2,stoch3}. The correct impact parameter treatment
of the pomeron loops was obtained in \cite{stoch2,stoch3}, whereas
the Langevin equation of \cite{stoch1} may be considered only as a 
some effective model of a correct Langevin equation.
Indeed, in \cite{stoch1} the semiclassical
treatment of large impact parameter behavior of the amplitude
was used,
when the sizes of the interacting dipoles are neglected in comparison to the
impact parameter of the problem, that excludes a correct treatment
of the pomeron's loops.
Therefore, talking about the comparison of the 
Langevin equations obtained in the framework
of QCD-RFT approach with the Langevin equations of the dipole approach, 
we will have in mind mostly the comparison of the
results of this paper with the
results of \cite{stoch2}. Nevertheless,
we also will give the derivation of results of \cite{stoch1}
in the framework of  QCD-RFT . It must be noticed here, that
the equivalence of  QCD-RFT and CGC approaches 
on the semiclassical level was considered as well in the framework of
a generating functional approach and BFKL pomeron calculus of
\cite{LevLu1,lev1}, where the problem of mutual
description of QCD-RFT and CGC was formulated and resolved
on the semiclassical level.
But, as it was mentioned above, in our paper, in spite of the papers
\cite{stoch1,lev1}, the Langevin equation will be obtained 
with proper consideration of an whole impact parameter structure of the theory,
where as a starting point the QCD-RFT approach will be used.

  The paper is organized as follows. In the next section we will introduce
and consider the main construction blocks of the QCD-RFT approach
and clarify they relations with such physically relevant quantity
as unintegrated gluon density. In the Section 3 we will obtain the 
Langevin equation for the theory with only triple pomeron vertexes of splitting
and merging. In the Section 4 we will introduce a "toy" 
four pomeron interaction vertex and will consider the possibility to write the 
Langevin equation for the theory with such additional vertex. The 
Section 5 will contain the comparison between the Langevin equations
obtained in s-chanel dipole model and t-channel QCD RFT model.
The Section 6 is a discussion of obtained results and 
conclusion of the paper. 

\section{Effective field theory of interacting pomerons}
 
\begin{figure}[tbp]
\begin{center}
\psfig{file=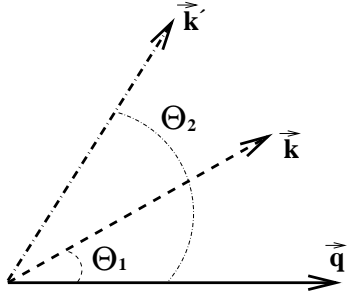,width=45mm} 
\end{center} 
\caption{\it The relative positions of the angles for the 
$\vec{k}$, $\vec{k^{'}}$ and $\vec{q}$ vectors.}
\label{Res1}
\end{figure}

 In this section we will formulate the main results concerning the 
pomeron effective theory considered in 
\cite{braun1,braun2,braun3,braun4}. 
A first ingredient, which we will need in the further 
calculations, is a action of the pomeron effective theory:
\beq\label{EFT1}
S\,=\,S_0\,+\,S_I\,,
\eeq
where $S_0$ and $S_I$ are the free and interacting
parts of the action correspondingly :
\beq\label{EFT2}
\,S_0\,=\,\int\,dy\,dy^{'}\,d^{2}r_{1}\,d^{2}r_{2}\,d^{2}r^{'}_{1}\,
d^{2}r^{'}_{2}\,\Php(y,r_{1},r_{2})\,
G^{-1}_{y-y^{'}}(r_{1},r_{2}|r^{'}_{1},r^{'}_{2})\,
\Ph(y,r^{'}_{1},r^{'}_{2})\,,
\eeq 
and
\beq\label{EFT3}
S_I \,=\,\frac{2\,\as^2\,N_c}{\pi}\,
\int\,dy\,\int\,\frac{d^{2}r_{1}\,d^{2}r_{2}\,d^{2}r_{3}}{\rad\,\radd\,\raddd}\,
\,(L_{13}\,\Ph(y,r_{1},r_{3}))\,\Php(y,r_{1},r_{2})\,\Php(y,r_{2},r_{3})\,
\,+\,
\eeq
$$
\,+\,
\,\frac{2\,\as^2\,N_c}{\pi}\,
\int\,dy\,\int\,\frac{d^{2}r_{1}\,d^{2}r_{2}\,d^{2}r_{3}}{\rad\,\radd\,\raddd}\,
\,(L_{13}\,\Php(y,r_{1},r_{3}))\,\Ph(y,r_{1},r_{2})\,\Ph(y,r_{2},r_{3})\,\,.
$$
In comparison with the action of \cite{braun3} we omitted here
the part of the action responsible for the interacting of 
pomerons with the target and projectile, the
source terms, because for our following calculations these terms are not important.
In the following derivations we also will not 
especially underline the fact that such quantities as
$r_i$ and/or $k_i$ always denote the two dimensional 
vectors, it will be denoted as vectors only in 
 the cases when precise definition of the vector structure
of $r_i$ and/or $k_i$ will be needed.
The part of the action, given by expression \eq{EFT3},
reproduces the triple pomeron vertex 
in the large $N_c$ limit with the use of an operator $L_{13}$:
\beq\label{EFT4}
L_{13}\,=\,r^{4}_{13}\,p^{2}_{1}\,p^{2}_{3} \,=\,r^{4}_{13}\,
\nabla^{2}_{1}\,\nabla^{2}_{3} \,\,.
\eeq
The propagator of the theory, 
$G^{-1}_{y-y^{'}}(r_{1},r_{2}|r^{'}_{1},r^{'}_{2})\,$, 
is defined throw the BFKL Hamiltonian \cite{Ham1,Ham2} :
\beq\label{EFT5}
G^{-1}_{y-y^{'}}(r_{1},r_{2}|r^{'}_{1},r^{'}_{2})\,=\,
\Le\,\nabla^{2}_{2}\,\nabla^{2}_{1}\,
\Le\,\frac{\D}{\D\,y}+H(r_{1},r_{2})\Ra\,\Ra\,
\delta^{2}(r_{1}\,-\,r^{'}_{1})\,
\delta^{2}(r_{2}\,-\,r^{'}_{2})\,
\delta(y-y^{'})\,\,.
\eeq
Now, let us consider only the triple pomeron interaction terms
of the action and let us
make a well known change of the variables in the action
\beq\label{EFT6}
\phi(r_{1},r_{2})\,\rightarrow\,
\frac{\Ph(r_{1},r_{2})}{\rad}\,\,\,,
\php(r_{1},r_{2})\,\rightarrow\,
\frac{\Php(r_{1},r_{2})}{\rad}\,,
\eeq
obtaining
\beq\label{EFT7}
S_I \, =\,\,\frac{2\,\as^2\,N_c}{\pi}\,
\int\,dy\,\int\,d^{2}r_{1}\,d^{2}r_{2}\,d^{2}r_{3}\,
\,(\frac{L_{13}}{\raddd}\,(\raddd \phi(y,r_{1},r_{3})))\,
\php(y,r_{1},r_{2})\,\php(y,r_{2},r_{3})\,
\,+\,
\eeq
$$
\,+\,
\,\frac{2\,\as^2\,N_c}{\pi}\,
\int\,dy\,\int\,d^{2}r_{1}\,d^{2}r_{2}\,d^{2}r_{3}\,
\,(\frac{L_{13}}{\raddd}\,(\raddd  \php(y,r_{1},r_{3})))\,
\phi(y,r_{1},r_{2})\,\phi(y,r_{2},r_{3})\,\,.
$$
In order to relate the amplitude with the 
unintegrated gluon density , as a next step, we perform Fourier transform:
\beq\label{EFT8}
\phi(y,r_{1},r_{2})\,=\,
\int\,\frac{d^2\,k_1\,d^2\,q_1\,}{(2\,\pi)^2}\,
e^{-i\,r_{1}\,k_1\,-i\,r_{2}\,(q_1\,-\,k_1)}\,
\tilde{\phi}(y,k_1,q_1-k_1)\,
\eeq
for the functions $\phi(y,r_{i},r_{j})$ in \eq{EFT7}.
In the terms of the functions $\phi(y,k_{i},q_{i}-k_{i})$ 
the triple pomeron term in the action will have the form:
\beq\label{EFT9}
S_I \,=\frac{2\,\as^2\,N_c}{\pi}
\int\,dy\int\,d^{2}k_{3}\,d^{2}q_{3}\,d^{2}k_{2}
\,(\hat{L}_{3}\,\tilde{\phi}(y,k_{3},q_{3}-k_{3}))
\tilde{\php}(y,q_{3}-k_{3},-k_{2})\,\tilde{\php}(y,k_{2},k_{3})
\,+\,
\eeq
$$
\,+\,
\,\frac{2\,\as^2\,N_c}{\pi}\,
\int\,dy\,\int\,d^{2}k_{3}\,d^{2}q_{3}\,d^{2}k_{2}\,
\,(\hat{L}_{3}\,\tilde{\php}(y,k_{3},q_{3}-k_{3}))\,
\tilde{\phi}(y,q_{3}-k_{3},-k_{2})\,\tilde{\phi}(y,k_{2},k_{3})\,
$$
where, 
\beq\label{EFT10}
\hat{L}_{3}\,=\,\nabla^{2}_{k_3}\,k^{2}_{3}\,
(q_{3}-k_{3})^{2}\,\nabla^{2}_{k_3}\,,
\eeq
see Appendix A for the detailed derivation,
and as usual, $q$ and $k$ here are two dimensional vectors.
Now we consider the case of interaction with zero  momentum 
transfer, when:
\beq\label{EFT11}
\int\,d^{2}\,q_{3}\,\tilde{\phi}(y,k_{3},q_{3}-k_{3})\,=\
\int\,d^{2}\,q_{3}\,\varphi(y,k_{3})\,\delta^{2}(q_{3})\,=\,
\varphi(y,k_{3})\,.
\eeq
In this case in \eq{EFT9} we have two delta functions
arisen from the substitutions:
\beqar\label{EFT12}
\tilde{\phi}(y,q_{3}-k_{3},-k_{2})\,& \rightarrow\,&
\varphi(y,q_{3}-k_{3})\,\delta^{2}(q_{3}\,-\,k_{3}\,-\,k_{2})\,\\
\tilde{\phi}(y,k_{2},k_{3})\,& \rightarrow\,&
\varphi(y,k_{3})\,\delta^{2}(k_{3}\,+\,k_{2})\,
\eeqar
that gives after the integration over $k_2$ and $q_3$
the triple pomeron vertex which is local in momentum  space:
\beq\label{EFT13}
S_I \,=\frac{2\,\as^2\,N_c}{\pi}
\int\,dy\int\,d^{2}k\,
\,(\,\nabla^{2}_{k}\,k^{4}\,
\,\nabla^{2}_{k}\,\varphi(y,k))\,
\varphi^{\dagger}(y,k)\,\varphi(y,k)^{\dagger}
\,+\,
\eeq
$$
\,+\,
\frac{2\,\as^2\,N_c}{\pi}
\int\,dy\int\,d^{2}k\,
\,(\,\nabla^{2}_{k}\,k^{4}\,
\,\nabla^{2}_{k}\,\varphi^{\dagger}(y,k))\,
\varphi(y,k)\,\varphi(y,k)\,.
$$
It is clear, that the function $\varphi(y,k)$ is a 
scattering amplitude for the case
when we omit a  momentum transfer, taking it equals zero,
i.e. for the case of forward scattering.
In the impact parameter representation this 
approximation corresponds to the semiclassical
approximation of the large impact parameter
limit, where the sizes of the interacting dipoles
are neglected in comparison to the whole impact parameter
of the problem.
This  $\varphi(y,k)$ function we may connect with  
the unintegrated gluon (parton) density function $f(k)$ 
with the help of the 
following expression \cite{braun2,kks}:
\beq\label{EFT14}
f(y,k) = {N_c \over 2 \pi^2} k^4 \nabla_k ^2 \varphi(y,k)\,,
\eeq
and, therefore, the physical meaning of the $\varphi(y,k)$
became to be clear:
through \eq{EFT14} $\varphi(y,k)$  defines 
the unintegrated gluon density function for the processes 
of forward scattering.
Basing on the  \eq{EFT9}-\eq{EFT10} and \eq{EFT13}-\eq{EFT14}
it is easy to generalize the expression \eq{EFT14}
for the case of non-zero momentum transfer:
\beq\label{EFT15}
\tilde{f}(y,k,q-k) = {N_c \over 2 \pi^2} k^2\,(q-k)^2 \nabla_k ^2\, 
\tilde{\phi}(y,k,q-k)\,,
\eeq
here $\tilde{\phi}(y,k,q-k)$ is the amplitude defined by \eq{EFT8}
and $\tilde{f}(y,k,q-k)$ is a generalized (skewed) gluon (parton) distribution
function. From \eq{EFT15} and \eq{EFT6} it is easy to see,
that our initial amplitude $\Ph(r_{1},r_{2})$
is simply Fourier transform of the $\tilde{f}(y,k,q-k)$ and vice versa:
\beq\label{EFT16} 
\frac{\tilde{f}(y,k,q-k)}{k^2\,(q-k)^2}\,=\, {N_c \over 2 \pi^2}\,
\int\,\frac{d^2\,r_1\,d^2\,r_2\,}{(2\,\pi)^2}\,
e^{i\,r_{1}\,k\,+i\,r_{2}\,(q\,-\,k)}\,
\Ph(r_{1},r_{2})\,
\eeq
that determines very clear and transparence meaning for the amplitude
$\Ph(r_{1},r_{2})$.
Considering the equations \eq{EFT14} and \eq{EFT15},
which  are the Poisson's type equations,
with the help of Green's function for the two dimensional
Poisson's equation, we can write the inverse relation between
$\tilde{f}(y,k,q-k)$ and $\tilde{\phi}(y,k,q-k)$  :
\beq\label{EFT17} 
\tilde{\phi}(y,k,q-k,\theta_{1})\,=  {2 \pi^2 \over N_c\,}
\int\,d^{2}k^{'}\,g(k,k^{'},q)\,
\frac{\tilde{f}(y,k^{'},q-k^{'},\theta_{2})}{k^{'2}\,(q-k^{'})^2}
\,=\,
\eeq
$$
= {2 \pi^2 \over N_c\,}\int\,\frac{d\theta_{2}}{4\,\pi}
\log\Le\frac{k^2\,-\,2\,k\,k^{'}
Cos(\theta_{1}-\theta_{2})+k^{'2}}{k^2}
\Ra\int\,k^{'}dk^{'}\,
\frac{\tilde{f}(y,k^{'},q-k^{'},\theta_{2})}{(k^{'})^{2}\,(q-k^{'})^2}\,.
$$
The relative positions of the angles $\theta_{1}$
and $\theta_{2}$ are denoted in the Fig.~\ref{Res1}.
Now it is easy to obtain the inverse expression for the \eq{EFT14}:
\beq\label{EFT18}
\varphi(y,k)\,=\,\int\,d^{2}\,q\,\tilde{\phi}(y,k,q-k,\theta_{1}=0)\,
\delta^{2}(q)\,=\,
\eeq
$$
\,=\,{2 \pi^2 \over N_c\,}
\int\,k^{'}dk^{'}\,d^{2}\,q\,
\frac{f(y,k^{'})}{k^{'2}\,(q-k^{'})^2}\,\delta^{2}(q)\,
\int\,\frac{d\theta_{2}}{4\,\pi}
\log\Le\frac{k^2\,-\,2\,k\,k^{'}
Cos(\theta_{2})+k^{'2}}{k^2}\Ra\,=\,
$$
$$
\,=\,{2 \pi^2 \over N_c\,}\,
\int_{0}^{k^2}\,k^{'}dk^{'}\,\frac{f(y,k^{'})}{k^{'4}}\,
\frac{2\pi\,\log(k^2)}{4\,\pi}\,+\,
{2 \pi^2 \over N_c\,}\,
\int_{k^2}^{\infty}\,k^{'}dk^{'}\,\frac{f(y,k^{'})}{k^{'4}}\,
\frac{2\pi\,\log(k^{'2})}{4\,\pi}\,-\,
$$
$$
\,-\,{2\pi^2 \over N_c\,}\,
\int_{0}^{\infty}\,k^{'}dk^{'}\,\frac{f(y,k^{'})}{k^{'4}}\,
\frac{2\pi\,\log(k^2)}{4\,\pi}\,=\,{\pi^2 \over 2\,N_c\,}\,
\int_{k^2}^{\infty}\,dk^{'2}\,\frac{f(y,k^{'})}{k^{'4}}\,
\log(\frac{k^{'2}}{k^2})\,,
$$
see also \cite{kks}. With the help of \eq{EFT13} and \eq{EFT18} 
we could obtain the expression for the triple pomeron vertex in 
terms of unintegrated gluon density, see \cite{kks,bond1} and Appendix B.
In general, using  \eq{EFT7} and \eq{EFT17}, it must be also 
possible to obtain the Bartels triple pomeron vertex  
written in the terms  of the skewed unintegrated
gluon densities $\tilde{f}(y,k,q-k)$ , see \cite{vert1},
but we do not consider this task
in this paper. So, in the further consideration, using the  
amplitudes $\Ph(y,r_{i},r_{j})$ or $\phi(y,r_{i},r_{j})$
defined in the transverse position space  we will always remember,
that these quantities are related to the generalized (skewed) 
gluon density function $\tilde{f}(y,k,q-k)$ in the momentum space.

\section{Langevin equation in the theory with the triple pomeron vertex}

\subsection{Langevin equation 
for the $\Ph(y,r_{i},r_{j})$ field in the transverse position space}

  In order to introduce an auxiliary field
in the theory we come back to the 
particular part of the action from the  \eq{EFT3}:

\beq\label{Lan1}
S_{\Ph}\,=\,\frac{2\,\as^2\,N_c}{\pi}\,
\int\,dy\,\int\,\frac{d^{2}r_{1}\,d^{2}r_{2}\,d^{2}r_{3}}{\rad\,\radd\,\raddd}\,
\,(L_{13}\,\Ph(y,r_{1},r_{3}))\,\Php(y,r_{1},r_{2})\,\Php(y,r_{2},r_{3})\,
\eeq
and we rewrite this expression in the following form:
\beq\label{Lan2}
S_{\Ph}\,=\,\frac{1}{2}\,
\int\,dy\,\frac{d^{2}\rho_{1}\,d^{2}\rho_{2}}{\rho^{4}_{12}}\,
\Php(y,\rho_{1},\rho_{2})\,
\int\,dy^{'}\,\frac{d^{2}\rho_{1}^{'}\,d^{2}\rho_{2}^{'}}
{\rho^{4}_{1^{'}2^{'}}}\,\Php(y^{'},\rho_{2}^{'},\rho_{1}^{'})\,
\eeq
$$
\Le\,\frac{4\as^2\,N_c}{\pi}\,
(L_{11^{'}}\,\Ph(y,\rho_{1},\rho_{1}^{'}))\,
\frac{\rho^{4}_{12}\,\rho^{4}_{1^{'}2^{'}}}{\rho^{2}_{12}\,
\rho^{2}_{1^{'}2}\,\rho^{2}_{11^{'}}}\,\delta(y-y^{'})\,
\delta^{2}(\rho_{2}-\rho_{2}^{'})\Ra\,.
$$
Now we introduce on the scene an auxiliary field through the 
Gaussian integration over the
auxiliary field $\psi$ :
\[
e^{S_{\Ph}}=\,
N\int D[\psi] exp\left\{-\frac{1}{2}
\int\,dy\,\frac{d^{2}\rho_{1}\,d^{2}\rho_{2}}{\rho^{4}_{12}}
\psi(y,\rho_{1},\rho_{2})
\Le\,\frac{4\,\as^2\,N_c}{\pi}\,
(L_{11^{'}}\,\Ph(y,\rho_{1},\rho_{1}^{'}))
\frac{\rho^{4}_{12}\,\rho^{4}_{1^{'}2^{'}}}{\rho^{2}_{12}
\rho^{2}_{1^{'}2}\,\rho^{2}_{11^{'}}}\Ra^{-1}\right.
\] 
\beq
\left.
\delta(y-y^{'})\,\delta^{2}(\rho_{2}-\rho_{2}^{'})\,
\int\,dy^{'}\,\frac{d^{2}\rho_{1}^{'}\,d^{2}\rho_{2}^{'}}
{\rho^{4}_{1^{'}2^{'}}}\,\psi(y^{'},\rho_{2}^{'},\rho_{1}^{'})\,
-\,\int\,dy\,\frac{d^{2}\rho_{1}\,d^{2}\rho_{2}}{\rho^{4}_{12}}\,
\psi(y,\rho_{1},\rho_{2})\,\Php(y,\rho_{1},\rho_{2})\,
\right\}\,,
\eeq\label{Lan3}
where a distribution functional for
$\psi$ has the following form:
\[
W[\psi]\,=\,
N\,exp\left\{-\frac{1}{2}
\int\,dy\,\frac{d^{2}\rho_{1}\,d^{2}\rho_{2}}{\rho^{4}_{12}}
\psi(y,\rho_{1},\rho_{2})
\Le\,\frac{4\,\as^2\,N_c}{\pi}\,
(L_{11^{'}}\,\Ph(y,\rho_{1},\rho_{1}^{'}))
\frac{\rho^{4}_{12}\,\rho^{4}_{1^{'}2^{'}}}{\rho^{2}_{12}
\rho^{2}_{1^{'}2}\,\rho^{2}_{11^{'}}}\Ra^{-1}\right.
\] 
\beq\label{Lan10}
\left.
\delta(y-y^{'})\,\delta^{2}(\rho_{2}-\rho_{2}^{'})\,
\int\,dy^{'}\,\frac{d^{2}\rho_{1}^{'}\,d^{2}\rho_{2}^{'}}
{\rho^{4}_{1^{'}2^{'}}}\,\psi(y^{'},\rho_{1}^{'},\rho_{2}^{'})\,
\right\}\,,
\eeq
with
\beq\label{Lan11}
N\,=\,\Le\int D[\psi]\,W[\psi]\,\Ra^{-1}.
\eeq
Let us define a new part of the action of the theory:
\beq\label{Lan4}
S_{Aux}\,=\,
-\frac{1}{2}\,
\int\,dy\,\frac{d^{2}\rho_{1}\,d^{2}\rho_{2}}{\rho^{4}_{12}}\,
\psi(y,\rho_{1},\rho_{2})\,
\Le\,\frac{\as^2\,N_c}{2\pi}\,
(L_{11^{'}}\,\Ph(y,\rho_{1},\rho_{1}^{'}))\,
\frac{\rho^{4}_{12}\,\rho^{4}_{1^{'}2^{'}}}{\rho^{2}_{12}\,
\rho^{2}_{1^{'}2}\,\rho^{2}_{11^{'}}}\,\Ra^{-1}\,
\eeq
$$
\delta(y-y^{'})\,\delta^{2}(\rho_{2}-\rho_{2}^{'})\,
\int\,dy^{'}\,\frac{d^{2}\rho_{1}^{'}\,d^{2}\rho_{2}^{'}}
{\rho^{4}_{1^{'}2^{'}}}\,\psi(y^{'},\rho_{2}^{'},\rho_{1}^{'})\,,
$$
and rewrite whole action in the following form:
\beq\label{Lan5}
S\,=\,S_0\,+\,S_{Aux}\,+\,S_{\Php}\,-\,
\int\,dy\,\frac{d^{2}\rho_{1}\,d^{2}\rho_{2}}{\rho^{4}_{12}}\,
\psi(y,\rho_{1},\rho_{2})\,\Php(y,\rho_{1},\rho_{2})\,,
\eeq
where as a $S_{\Php}$ we denoted
a second part of the action $S_I$ from \eq{EFT3}.
Writing the equation of motion  for the 
$\Ph(y,\rho_{i},\rho_{j})$ field
\beq\label{Lan6}
\frac{\delta\,S}{\delta\,\Php(y,\rho_{1},\rho_{2})}\,=\,0\,
\eeq
we see, that we obtain the equation for the field $\Ph$
which does not depend on the field $\Php$ 
but rather on the field $\psi$:
\beq\label{Lan7}
G^{-1}\,\Ph(y,\rho_{1},\rho_{3})\,+\,\frac{2\,\as^2\,N_c}{\pi}\,
\int\,\frac{d^{2}\rho_{2}}
{\rho^{2}_{12}\,\rho^{2}_{23}\,\rho^{2}_{31}\,}\,
\Ph(y,\rho_{1},\rho_{2})\,\Ph(y,\rho_{2},\rho_{3})\,
\,(L_{13})\,-\frac{\psi(y,\rho_{1},\rho_{3})}{\rho^{4}_{13}}\,
=\,0\,,
\eeq
or
\beq\label{Lan8}
\Le  \frac{\D}{\D\,y}+H(\rho_{1},\rho_{3})\Ra
\,\Ph(y,\rho_{1},\rho_{3})\,+\,
\eeq
$$
\frac{2\,\as^2\,N_c}{\pi}\,
\int\,\frac{d^{2}\rho_{2}\,\rho^{2}_{31}}
{\rho^{2}_{12}\,\rho^{2}_{23}}\,
\,\Ph(y,\rho_{1},\rho_{2})\,\Ph(y,\rho_{2},\rho_{3})\,
-\,(L_{13})^{-1}\,\psi(y,\rho_{1},\rho_{3})\,
=\,0\,.
$$
The field $\psi$ in \eq{Lan8} may be considered as a noise
field with the following properties:
\beqar
<\,\psi(y,\rho_{1},\rho_{2})\,>\,& = & \,0\,\,;\\
\label{Lan9}\,
<\,\psi(y,\rho_{1},\rho_{2})\,,
\psi(y^{'},\rho_{2}^{'},\rho_{1}^{'})\,>\,& = & \,
\frac{4\as^2\,N_c}{\pi}\,
\frac{\rho^{2}_{12}\,\rho^{2}_{1^{'}2^{'}}}{\rho^{2}_{11^{'}}}\,
(L_{11^{'}}\,\Ph(y,\rho_{1},\rho_{1}^{'}))\,
\delta(y-y^{'})\,
\delta^{2}(\rho_{2}-\rho_{2}^{'})\,
\eeqar
that follows from the form of the distribution functional \eq{Lan10} for
the $\psi$.
The last term in \eq{Lan8} may be rewritten with the help of
the following properties of initial Green's function,
i.e. Green's function at zero rapidity. Indeed,
let us consider this Green's function 
$G_{0}(\rho_{1},\rho_{3}|\rho_{1^{'}},\rho_{3^{'}})$,
which has a form, see \cite{Ham2},
\beq\label{Lan12}
G_{0}(\rho_{1},\rho_{3}|\rho_{1^{'}},\rho_{3^{'}})\,=\,
\,\pi^{2}\,\ln\,\frac
{\rho_{11^{'}}^{2} \rho_{33^{'}}^{2}}
{\rho_{13^{'}}^{2} \rho_{1^{'}3}^{2}}\,
\ln\,\frac{\rho_{11^{'}}^{2} \rho_{33^{'}}^{2}}
{\rho_{13}^{2} \rho_{1^{'}3^{'}}^{2}}\,
\eeq
and satisfies
\beq\label{Lan13}
(\rho_{13})^{-4}\,L_{13}\Le\,
G_{0}(\rho_{1},\rho_{3}|\rho_{1^{'}},\rho_{3^{'}})\Ra\,=
\,(2\,\pi)^{4}\,\delta^{2}(\rho_{1}-\rho_{1^{'}})\,
\delta^{2}(\rho_{3}-\rho_{3^{'}})\,.
\eeq
Using \eq{Lan13} we obtain an operator identity:
\beq\label{Lan14}
(\nabla_{\rho_{1}}^{2})^{-1}\,(\nabla_{\rho_{3}}^{2})^{-1}\,=
\,\int\,\frac{d^{2}\,\rho_{1^{'}}\,d^{2}\,\rho_{3^{'}}}{(2\,\pi)^{4}}\,
G_{0}(\rho_{1},\rho_{3}|\rho_{1^{'}},\rho_{3^{'}})\,.
\eeq
With the help of \eq{Lan14} we rewrite \eq{Lan8} in the following form:
\beq\label{Lan15}
\Le  \frac{\D}{\D\,y}+H(\rho_{1},\rho_{3})\Ra
\,\Ph(y,\rho_{1},\rho_{3})\,+\,
\eeq
$$
\frac{2\,\as^2\,N_c}{\pi}
\int\,\frac{d^{2}\rho_{2}\,\rho^{2}_{31}}
{\rho^{2}_{12}\,\rho^{2}_{23}}
\Ph(y,\rho_{1},\rho_{2})\Ph(y,\rho_{2},\rho_{3})\,
-\int\,\frac{d^{2}\,\rho_{1^{'}}\,d^{2}\rho_{3^{'}}}{(2\,\pi)^{4}}
G_{0}(\rho_{1},\rho_{3}|\rho_{1^{'}},\rho_{3^{'}})\,
\frac{\psi(y,\rho_{1},\rho_{3})}{\rho_{13}^{4}}\,
=\,0\,.
$$
with the correlator for the function $\psi(y,\rho_{1},\rho_{3})$  
given by  \eq{Lan9}. Both \eq{Lan8} and \eq{Lan15}
with the auto correlator for the noise field \eq{Lan9}
are the required Langevin equations of the QCD-RFT approach,
valid for any value of the impact parameter of the problem.
These equations
are pretty complicated, therefore, let's try to simplify them
using the conformal basis representation for the pomeron
fields and assuming special properties of the theory at
high energy limit.

 The following property
of the field $\Ph(y,\rho_{i},\rho_{j})$ at high energy limit
of the theory could help us to simplify the Langevin equation.
Let us expand the $\Ph(y,\rho_{1},\rho_{2})$
field on the conformal basis formed by the functions
$E_{\mu(n,\nu)\,,\rho_ 0}(\rho_{1},\rho_{2})$ , \cite{lip1}:
\beq\label{Lan16}
E_{\mu(n,\nu)\,,\rho_ 0}(\rho_{1},\rho_{2})\,=\,
\Le\,\frac{\rho_{1}}{\rho_{10}\,\rho_{20}}
\Ra\,^{\frac{1-n}{2}+i\nu}\,
\Le\,\frac{\rho_{12}^{*}}{\rho_{10}^{*}\,\rho_{20}^{*}}
\Ra\,^{\frac{1+n}{2}+i\nu}\,
\eeq
and
\beq\label{Lan17}
L_{13}\,E_{\mu(n,\nu)\,,\rho_ 0}(\rho_{1},\rho_{2})\,=\,
\lambda_{\mu(n,\nu)}^{-1}\,E_{\mu(n,\nu)\,,\rho_ 0}(\rho_{1},\rho_{2})\,,
\eeq
with
\beq\label{Lan18}
\lambda_{\mu(n,\nu)}\,=\,\lambda_{\mu}=\,
\frac{1}{\Le\,(n+1)^2\,+4\nu^2\Ra\,
\Le\,(n-1)^2\,+4\nu^2\Ra}\,,
\eeq
where we used the same notations as in \cite{braun3}. 
This expansion has the form
\beq\label{Lan19}
\Ph(y,\rho_{1},\rho_{2})\,=\,\sum_{\mu}\,
E_{\mu(n,\nu)\,,\rho_ 0}(\rho_{1},\rho_{2})\,
\Ph_ {\mu}(y)\,=\,\sum_{\mu}\,
E_{\mu}(\rho_{1},\rho_{2})\,\Ph_ {\mu}(y)\,
\eeq
where
\beq\label{Lan20}
\Ph_ {\mu}(y)\,=\,\int\,
\frac{d^{2}\,\rho_{1}\,d^{2}\rho_{2}}{\rho_{12}^{4}}\,
E_{\mu}^{*}(\rho_{1},\rho_{2})\,
\Ph(y,\rho_{1},\rho_{2})\,,
\eeq
see \cite{lip1}. In \eq{Lan19} and in the following
expressions the notation of the conformal summation
always means 
\beq\label{Lan21}
\sum_{\mu}\,=\,\sum_{n=-\infty}^{\infty}\,
\int\,d\nu\,\frac{\nu^2\,+\,\frac{n^2}{4}}{\pi^4}\,
\int\,d^{2}\rho_{0}\,.
\eeq
At high energy limit , as it obtains for the  BFKL equation
 at high energy limit, see for example \cite{bond3},
we can assume that the main contribution in the
sum in \eq{Lan19} comes from the minimal conformal weight,
namely when $n=0$ and $\nu=0$. In this case 
we  have
\beq\label{Lan22}
L_{12}\Ph(y,\rho_{1},\rho_{2})\,=\,L_{12}\,
\sum_{\mu}\,E_{\mu}(\rho_{1},\rho_{2})\,\Ph_ {\mu}(y)\,
=\,\sum_{\mu}\,\lambda_{\mu}^{-1}\,
E_{\mu}(\rho_{1},\rho_{2})\,\Ph_ {\mu}(y)\,\simeq\,
\eeq
$$
\,\simeq\,\sum_{\mu}\,
E_{\mu}(\rho_{1},\rho_{2})\,\Ph_ {\mu}(y)\,=\,
\Ph(y,\rho_{1},\rho_{2})\,.
$$
Here we used the fact, that in the high energy limit when $n=0$ and $\nu=0$
we have $\lambda_{\mu(n=0,\nu=0)}^{-1}\,=1$.
The same approximation was used, for example,  in \cite{lev1}. 
It must be clear, 
that
this high energy approximation, which is based on the behavior of the BFKL amplitude, 
i.e. single pomeron, 
may be not correct in general in the theory of interacting pomerons.  
Now, supposing that the \eq{Lan22} is valid at high energy limit, 
we rewrite the  \eq{Lan15}
and \eq{Lan9} in the following form:
\beq\label{Lan23}
\Le  \frac{\D}{\D\,y}+H(\rho_{1},\rho_{3})\Ra
\,\Ph(y,\rho_{1},\rho_{3})\,+\,
\eeq
$$
\frac{2\,\as^2\,N_c}{\pi}
\int\,\frac{d^{2}\rho_{2}\,\rho^{2}_{31}}
{\rho^{2}_{12}\,\rho^{2}_{23}}
\Ph(y,\rho_{1},\rho_{2})\Ph(y,\rho_{2},\rho_{3})\,
-\int\,\frac{d^{2}\,\rho_{1^{'}}\,d^{2}\rho_{3^{'}}}{(2\,\pi)^{4}}
G_{0}(\rho_{1},\rho_{3}|\rho_{1^{'}},\rho_{3^{'}})\,
\psi(y,\rho_{1},\rho_{3})\,
=\,0\,.
$$
\beqar
<\,\psi(y,\rho_{1},\rho_{2})\,>\,& = & \,0\,\,;\\
\label{Lan24}
<\,\psi(y,\rho_{1},\rho_{2})\,,
\psi(y^{'},\rho_{2}^{'},\rho_{1}^{'})\,>\,& = & \,
\frac{4\as^2\,N_c}{\pi\Le\,
\rho^{2}_{12}\,\rho^{2}_{1^{'}2^{'}}\,\rho^{2}_{11^{'}}
\Ra}\,
\,\Ph(y,\rho_{1},\rho_{1}^{'})\,
\delta(y-y^{'})\,
\delta^{2}(\rho_{2}-\rho_{2}^{'})\,,
\eeqar
were we made the following substitution  
$\psi(y,\rho_{1},\rho_{2})\rightarrow\psi(y,\rho_{1},\rho_{2})/\rho_{12}^{4}$.
The equations \eq{Lan23}-\eq{Lan24} may be considered
as a high energy approximation to the correct Langevin equation
\eq{Lan8} with the auto correlator \eq{Lan9}.

\subsection{Langevin equation 
for the $\varphi(y,k)$ field in the momentum space}

 Let us now consider the part of the action
given by \eq{EFT13} for the case of the forward scattering 
and formulated in the momentum space:
\beq\label{Phi1}
S_{\varphi} \,=\frac{2\,\as^2\,N_c}{\pi}
\int\,dy\int\,d^{2}k\,
\,(\,\nabla^{2}_{k}\,k^{4}\,
\,\nabla^{2}_{k}\,\varphi(y,k))\,
\varphi^{\dagger}(y,k)\,\varphi(y,k)^{\dagger}\,.
\eeq
As it was mentioned previously, a dual expression
of the $S_{\varphi}$ part of the action
in the coordinate space will represent
a semiclassical approximation of the initial 
action \eq{EFT1}-\eq{EFT2}, and , therefore,
the results of this subsection are related 
to the Langevin equation of \cite{stoch1}.
So, performing the same transformations as in the previous section, 
we rewrite the $S_{\varphi}$ part of the action in the following form:
\beq\label{Phi2}
S_{\varphi}=\frac{1}{2}
\int dy\int\,d^{2}k\varphi^{\dagger}(y,k)
\Le\frac{4\as^2\,N_c}{\pi}(\nabla^{2}_{k}\,k^{4}
\nabla^{2}_{k}\varphi(y,k))
\delta(y-y^{'})\delta^{2}(k-k^{'})\Ra
\int\,dy^{'}\int\,d^{2}k^{'}
\varphi^{\dagger}(y^{'},k^{'})\,.
\eeq
Introducing an auxiliary field $\psi(y,k)$
we write for the $e^{S_{\varphi}}$:
\[
e^{S_{\varphi}}=\,
N\int D[\psi] exp\left\{-\frac{1}{2}\,
\int dy\,d^{2}k\,\psi(y,k)\,
\Le\frac{4\as^2\,N_c}{\pi}(\nabla^{2}_{k}\,k^{4}
\nabla^{2}_{k}\varphi(y,k))\,\Ra^{-1}\,\right.
\] 
\beq
\left.
\delta(y-y^{'})\delta^{2}(k-k^{'})\,
\int\,dy^{'}\,d^{2}k^{'}\,\psi(y^{'},k^{'})\,-\,
\int dy\,d^{2}k\,\psi(y,k)\,\varphi^{\dagger}(y,k)\,
\right\}\,.
\eeq\label{Phi3}
Reabsorbing the $\Le\frac{4\as^2\,N_c}{\pi}(\nabla^{2}_{k}\,k^{4}
\nabla^{2}_{k}\varphi(y,k))\,\Ra^{-1/2}$ factor in the definition
of the field $\psi$, we obtain:
\[
e^{S_{\varphi}}=\,
N\int D[\psi] exp\left\{-\frac{1}{2}\,
\int dy\,d^{2}k\,\psi(y,k)\,
\int\,dy^{'}\,d^{2}k^{'}\,\psi(y^{'},k^{'})\,
\,\delta(y-y^{'})\delta^{2}(k-k^{'})\,\right.
\] 
\beq
\left.
\,-\,
\int dy\,d^{2}k\,\psi(y,k)\,\varphi^{\dagger}(y,k)\,
\sqrt{\frac{4\as^2\,N_c}{\pi}(\nabla^{2}_{k}\,k^{4}
\nabla^{2}_{k}\varphi(y,k))}
\right\}\,.
\eeq\label{Phi4}
As in the previous case, N here is
\beq\label{Phi5}
N=\Le\,\int D[\psi] exp\left\{-\frac{1}{2}\,
\int dy\,d^{2}k\,\psi(y,k)\,
\int\,dy^{'}\,d^{2}k^{'}\,\psi(y^{'},k^{'})\,
\,\delta(y-y^{'})\delta^{2}(k-k^{'})\,\right\}\Ra^{-1}\,.
\eeq
The whole action for the $\varphi(y,k)$ and
$\varphi^{\dagger}(y,k)$ fields now takes the form:
\beq\label{Phi6}
S=S_0\,+\,S_{Aux}\,+\,S_{\varphi^{\dagger}}\,-\,
\int dy\,d^{2}k\,\psi(y,k)\,\varphi^{\dagger}(y,k)\,
\sqrt{\frac{4\as^2\,N_c}{\pi}(\nabla^{2}_{k}\,k^{4}
\nabla^{2}_{k}\varphi(y,k))}\,.
\eeq
where
\beq\label{Phi7}
\,S_{Aux}\,=\,-\frac{1}{2}\,\int dy\,\int d^{2}k\,\psi^{2}(y,k)\,.
\eeq
Writing an equation of motion for the field $\varphi(y,k)$
\beq\label{Phi8}
\frac{\delta S}{\delta \varphi^{\dagger}(y,k)}\,=\,0\,
\eeq
we obtain equation similar to \eq{Lan7}:
\beq\label{Phi9}
\hat{L}_{k}\,\Le\,\frac{\D}{\D\,y}+H(k)\Ra\,\varphi(y,k)\,+\,
\frac{2\,\as^2\,N_c}{\pi}\,
\varphi(y,k)\,\hat{L}_{k}\,-\,\psi(y,k)\,
\sqrt{\frac{4\as^2\,N_c}{\pi}(\hat{L}_{k}\,\varphi(y,k))}\,=\,0\,,
\eeq
with the operator $\hat{L}_{k}$ from the \eq{EFT13}:
\beq\label{Phi10}
\hat{L}_{k}\,=\,\nabla^{2}_{k}\,k^{4}\,
\,\nabla^{2}_{k}\,.
\eeq
We rewrite \eq{Phi9}:
\beq\label{Phi11}
\Le\,\frac{\D}{\D\,y}+H(k)\Ra\,\varphi(y,k)\,+\,
\frac{2\,\as^2\,N_c}{\pi}\,
\varphi(y,k)\,-\,
\Le\hat{L}_{k}\Ra^{-1}\,\Le\psi(y,k)\,
\sqrt{\frac{4\as^2\,N_c}{\pi}(\hat{L}_{k}\,\varphi(y,k))}\Ra\,=\,0\,,
\eeq
where correlators for the auxiliary filed $\psi(y,k)$
have the form:
\beqar
<\,\psi(y,k)\,>\,& = & \,0\,\,;\\
<\,\psi(y,k)\,,\psi(y_1,k_1)\,>\,& = & \,
\delta(y\,-\,y_1)\,\delta^{2}(k\,-\,k_1)\,.
\eeqar\label{Phi12}
It is important to underline, that contrary to the
simplifications obtained for the equations \eq{Lan23}
and \eq{Lan24}
for the $\Ph(y,\rho_{i},\rho_{j})$ field
at high energy limit, here we cannot write that
$\hat{L}_{k}\,\varphi(y,k)=\varphi(y,k)$. Indeed, 
such condition leads to the very non physical restriction
on the unintegrated parton density function $f(y,k)$
which follows from the \eq{EFT14} and \eq{EFT18}:
\beq\label{Phi13}
\nabla^{2}_{k}\,f(y,k)\,=\,\frac{1}{4}\,
\int_{k^2}^{\infty}\,dk^{'2}\,\frac{f(y,k^{'})}{k^{'4}}\,
\log(\frac{k^{'2}}{k^2})\,.
\eeq
Of course, such identity can not be satisfied in general 
for arbitrary $f(y,k)$ function. More of that, due the 
use of semiclassical approximation in derivation of this result,
it is not clear at all, how such approximated Langevin equation
may be used as a guideline for the pomeron loops calculations.
Indeed, by definition, a loops calculation must correctly treat
the impact parameter structure of the theory , whereas the \eq{Phi11}
and \eq{Phi12} were obtained in the approximation
of the large impact parameter.

\section{Langevin equation 
for the theory with "toy" four pomeron vertex}

\subsection{"Toy" four pomeron vertex in the effective theory of the
interacting pomerons }

 The Lagrangian of the RFT-0 model for the $q$ and $p$ pomeron fields, 
which includes also a four pomeron vertex,
we could define in the following form: 
\beq\label{FP1}
L=q\,\dot{p}\,+\,\mu\,q \,p\,-\,\lambda\,q\,(q\,+\,p)\,p\,+\,
\lambda^{'}\,q^2\,p^2\,,
\eeq
see \cite{0dim1,0dim2,bond2},
where $\mu$ is a bare pomeron intercept, $\lambda$
is a vertex of triple pomeron interactions and $\lambda^{'}$
a four pomeron interaction vertex. For the case of "fine tuning"
of the vertexes, when $\frac{\mu}{\lambda}=\frac{\lambda}{\lambda^{'}}$,
that defines a "magic" value of the four pomeron vertex $\lambda^{'}$,
the Hamiltonian of the problem has 
a factorized form in the terms of $q$ and $p$ fields :
\beq\label{FP2}
-H=\,\mu\,(q\,-\,\frac{\lambda}{\mu}\,q^2)\,p\,-\,
\lambda\,(q\,-\,\frac{\lambda^{'}}{\lambda}\,q^2)\,p^2\,
\,=\,\mu\,(q\,-\,\frac{\lambda}{\mu}\,q^2)\,(\,p\,-\,
\frac{\lambda}{\mu}\,p^2\,)\,.
\eeq
We are not interesting in the further investigation of the RFT-0 model here,
see more details in \cite{bond2,lev1,ham0}, but as a guideline for
the derivation of our "toy" four pomeron vertex, we will take
the same as in \eq{FP2}
property of the factorizability of the Hamiltonian
for the case of "magic" value of the four pomeron vertex.

 We will begin from the free part
of the effective pomeron theory action, \eq{EFT2}, written
in the following form:
\beq\label{FP3}
S_0=\frac{1}{2}\,\int\,
\,dy\,\frac{d^{2}r_{1}\,d^{2}r_{3}}{r_{13}^{4}}\,\Le\,
\Php(y,r_{1},r_{3})\,
\frac{\D (L_{13}\Ph(y,r_{1},r_{3}))}{\D y}\,
-\,\frac{\D (L_{13}\Php(y,r_{1},r_{3}))}{\D y}\,
\Ph(y,r_{1},r_{3})\Ra\,+\,
\eeq
$$
+\,\frac{1}{2}\int\,
dy\frac{d^{2}r_{1}\,d^{2}r_{3}}{r_{13}^{4}}\Le\,
\Ph(y,r_{1},r_{3})
(L_{13}H(r_{1},r_{3})\,\Php(y,r_{1},r_{3}))+
(L_{13}H(r_{1},r_{3})\,\Ph(y,r_{1},r_{3}))
\Php(y,r_{1},r_{3})\Ra\,.
$$
Here $H(r_{1},r_{3})$ is a BFKL Hamiltonian, \cite{Ham1,Ham2,lip1}.
In order not to confuse this Hamiltonian
with general Hamiltonian of the problem $H$, the BFKL
Hamiltonian and operators related with BFKL Hamiltonian
will be always written with the arguments
of the BFKL Hamiltonian, namely in the form $H(r_{i},r_{j})$. 
Writing the Hamiltonian of the problem
in the conformal basis, formed by functions 
$E_{\mu}(r_{1},r_{3})$ of \eq{Lan16}, we will
omit in the further expressions the common integration factor
\beq\label{fac}
\int\,dy\int\frac{d^{2}r_{1}\,d^{2}r_{3}}{r_{13}^{4}}\,.
\eeq
The expression for the "free" part of the Hamiltonian
in the conformal basis is the following:
\beqar\label{FP4}
H_{0}\,& = &\,\frac{1}{2}\,\sum_{\mu}\,
E_{\mu}(r_{1},r_{3})\,\omega_{\mu}\,\lambda_{\mu}^{-1}\,
\Ph_{\mu}(y)\,\sum_{\nu}\,E_{\nu}^{*}(r_{1},r_{3})
\Php_{\nu}(y)\,+\,\\
\, & + & \,\frac{1}{2}\,\sum_{\mu}\,
E_{\mu}(r_{1},r_{3})\,\Ph_{\mu}(y)\,\sum_{\nu}\,
E_{\nu}^{*}(r_{1},r_{3})\,
\omega_{\nu}\,\lambda_{\nu}^{-1}\,\Php_{\nu}(y)\,,
\eeqar
where we used \eq{Lan17} and 
\beq\label{FP5}
H(r_{1},r_{3})\,E_{\mu}(r_{1},r_{3})\,=\,-\,\omega_{\mu}\,E_{\mu}(r_{1},r_{3})\,,
\eeq
with the eigenvalues $\omega_{\mu}$ for the eigenfunctions  
$E_{\mu}(r_{1},r_{2})\,$ of the BFKL Hamiltonian \cite{lip2}.
Before the definition of the "interacting" part of the Hamiltonian, let us
introduce functions
\beq\label{FP6}
\Psi(y,r_1,r_3)\,=\,\int\,
\frac{d^{2}r_{2}\,r_{13}^{2}}{r_{12}^{2}\,r_{23}^{2}}\,
\Ph(y,r_{1},r_{2})\,\Ph(y,r_{2},r_{3})\,=\,
\sum_{\mu}\,E_{\mu}(r_{1},r_{3})\,\Psi_{\mu}(y)\,,
\eeq
and
\beq\label{FP7}
\Psi^{\dagger}(y,r_1,r_3)\,=\,\int\,
\frac{d^{2}r_{2}\,r_{13}^{2}}{r_{12}^{2}\,r_{23}^{2}}\,
\Php(y,r_{1},r_{2})\,\Php(y,r_{2},r_{3})\,=\,
\sum_{\mu}\,E_{\mu}^{*}(r_{1},r_{3})\,\Psi_{\mu}^{\dagger}(y)\,.
\eeq
With the use of the  $\Psi(y,r_1,r_3)$ and $\Psi^{\dagger}(y,r_1,r_3)$
functions the "interacting" part of the Hamiltonian,
which corresponds to \eq{EFT3},
obtains the following form:
\beqar\label{FP8}
-H_{I}\,& = &\,\frac{2\,\as^2\,N_c}{\pi}\,
\sum_{\mu}\,E_{\mu}(r_{1},r_{3})\,\lambda_{\mu}^{-1}\,
\Ph_{\mu}(y)\,\sum_{\nu}\,E_{\nu}^{*}(r_{1},r_{3})\,
\Psi_{\nu}^{\dagger}(y)\,+\,\\
\,& + &\frac{2\,\as^2\,N_c}{\pi}\,
\sum_{\mu}\,E_{\mu}^{*}(r_{1},r_{3})\,\lambda_{\mu}^{-1}\,
\Php_{\mu}(y)\,\sum_{\nu}\,E_{\nu}(r_{1},r_{3})\,
\Psi_{\nu}(y)\,.
\eeqar
Let us now assume the following anzats 
for the action with the
"toy" four pomeron interaction vertex :
\beq\label{FP9}
S_{4P}\,= \,-\,C\,
\int\,dy\int\,
\frac{d^{2}r_{1}\,d^{2}r_{3}}{r_{13}^4}\,
\hat{F}_{4P}\Le\,\sum_{\mu}\,E_{\mu}(r_{1},r_{3})\,
\Psi_{\nu}(y)\,\sum_{\nu}\,E_{\nu}^{*}(r_{1},r_{3})\,
\Psi_{\nu}^{\dagger}(y)\,\Ra\,=\,
\eeq 
\[
\, =  \,
\int\,dy\int\,
\frac{d^{2}r_{1}\,d^{2}r_{3}}{r_{13}^4}\,
\Le\,
\sum_{\mu}\,(\hat{F}_{1}\,E_{\mu}(r_{1},r_{3}))\,
\Psi_{\nu}(y)\,\sum_{\nu}\,(\hat{F}_{2}\,E_{\nu}^{*}(r_{1},r_{3}))\,
\Psi_{\nu}^{\dagger}(y)\,+\,\right.
\] 
\[
\left.
\,+\,\sum_{\mu}\,(\hat{F}_{2}\,E_{\mu}(r_{1},r_{3}))\,
\Psi_{\nu}(y)\,\sum_{\nu}\,(\hat{F}_{1}\,E_{\nu}^{*}(r_{1},r_{3}))\,
\Psi_{\nu}^{\dagger}(y)\,\Ra\,=\,
\] 
\[
\, = \,
\int\,dy\int\,
\frac{d^{2}r_{1}\,d^{2}r_{3}}{r_{13}^4}\,
\Le\,
\sum_{\mu}\,f_{1\mu}\,E_{\mu}(r_{1},r_{3})\,
\Psi_{\nu}(y)\,\sum_{\nu}\,f_{2\nu}\,E_{\nu}^{*}(r_{1},r_{3})\,
\Psi_{\nu}^{\dagger}(y)\,+\,
\right.
\] 
$$
\left.
\,+\,\sum_{\mu}\,f_{2\mu}\,E_{\mu}(r_{1},r_{3})\,
\Psi_{\nu}(y)\,\sum_{\nu}\,f_{1\nu}\,E_{\nu}^{*}(r_{1},r_{3})\,
\Psi_{\nu}^{\dagger}(y)\,\Ra\,,
$$
where
we introduced some operators $\hat{F}_{1}$ and $\hat{F}_{2}$
such that
\beq\label{FP10}
\hat{F}_{i}\,E_{\mu}(r_{1},r_{3})\,=\,f_{i\mu}\,E_{\mu}(r_{1},r_{3})\,.
\eeq
Finally, again omitting the integration over 
$\int\,dy\int\,d^{2}r_{1}\,d^{2}r_{3}\,/\,r_{13}^{4}$,
we write the Hamiltonian which is
corresponding to this anzats:
\beqar\label{FP11}
H_{4P}\,& = &\,C\,
\sum_{\mu}\,f_{1\mu}\,E_{\mu}(r_{1},r_{3})\,
\Psi_{\nu}(y)\,\sum_{\nu}\,f_{2\nu}\,E_{\nu}^{*}(r_{1},r_{3})\,
\Psi_{\nu}^{\dagger}(y)\,+\,\\
\,& + & C\,\sum_{\mu}\,f_{2\mu}\,E_{\mu}(r_{1},r_{3})\,
\Psi_{\nu}(y)\,\sum_{\nu}\,f_{1\nu}\,E_{\nu}^{*}(r_{1},r_{3})\,
\Psi_{\nu}^{\dagger}(y)\,,
\eeqar
where the constant $C$  will be defined in the further derivation
of the Hamiltonian.
Now, collecting all terms \eq{FP4}, \eq{FP8} and \eq{FP11} together,
we obtain:
\beqar\label{FP12}
H & = & \frac{1}{2}\,\sum_{\mu}\,
E_{\mu}(r_{1},r_{3})\,\omega_{\mu}\,\lambda_{\mu}^{-1}\,
\Ph_{\mu}(y)\sum_{\nu}\,E_{\nu}^{*}(r_{1},r_{3})\Le
\Php_{\nu}(y)-\frac{2\as^2\,N_c}{\pi}
\omega_{\mu}^{-1}\,\Psi_{\nu}^{\dagger}(y)\,\Ra\,+\,\\
& + &\,\frac{1}{2}\,\sum_{\mu}\,
E_{\mu}^{*}(r_{1},r_{3})\,\omega_{\mu}\,\lambda_{\mu}^{-1}\,
\Php_{\mu}(y)\,\sum_{\nu}\,E_{\nu}(r_{1},r_{3})\Le
\Ph_{\nu}(y)-\frac{2\,\as^2\,N_c}{\pi}\,
\omega_{\mu}^{-1}\,\Psi_{\nu}(y)\,\Ra\,-\,\\
& - &\,\frac{\as^2\,N_c}{\pi}\sum_{\mu}\,
E_{\mu}(r_{1},r_{3})\,\Psi_{\mu}(y)\,
\sum_{\nu}\,E_{\nu}^{*}(r_{1},r_{3})\lambda_{\nu}^{-1}\Le
\Php_{\nu}(y)-\frac{C\pi}{\as^2\,N_c}\,
\lambda_{\nu}f_{1\mu}f_{2\nu}\,\Psi_{\nu}^{\dagger}(y)\,\Ra\,-\,\\
& - &\,\frac{\as^2\,N_c}{\pi}\sum_{\mu}\,
E_{\mu}^{*}(r_{1},r_{3})\,\Psi_{\mu}^{\dagger}(y)\,
\sum_{\nu}\,E_{\nu}(r_{1},r_{3})\lambda_{\nu}^{-1}\Le
\Ph_{\nu}(y)-\frac{C\pi}{\as^2\,N_c}\,
\lambda_{\nu}f_{1\mu}f_{2\nu}\,\Psi_{\nu}(y)\,\Ra\,.
\eeqar
Assuming for \eq{FP12} the same factorization property as for
the Hamiltonian of RFT-0, we find
the following values for the $C\,,f_{1\mu}\,,f_{2\nu}$:
\beq\label{FP13}
C\,=\,2\,\Le\frac{\as^2\,N_c}{\pi}\Ra^{2}\,,\,\,\,\,
f_{1\mu}\,=\omega_{\mu}^{-1}\,,\,\,\,\,
f_{2\nu}\,=\lambda_{\nu}^{-1}\,.
\eeq
Continuing the derivation, we obtain for the Hamiltonian:
\beqar\label{FP14}
H & = & \sum_{\mu}E_{\mu}(r_{1},r_{3})
\sum_{\nu}E_{\nu}^{*}(r_{1},r_{3})\Le
\omega_{\mu}\lambda_{\mu}^{-1}\frac{\Ph_{\mu}(y)}{2}-
\frac{\as^2\,N_c}{\pi}\Psi_{\mu}(y)\lambda_{\nu}^{-1}\Ra\,\\
&\,&
\Le\Php_{\nu}(y)-\frac{2\as^2\,N_c}{\pi}
\omega_{\mu}^{-1}\Psi_{\nu}^{\dagger}(y)\Ra\,+\,\\
&+& \sum_{\mu}E_{\mu}^{*}(r_{1},r_{3})
\sum_{\nu}E_{\nu}(r_{1},r_{3})\Le
\omega_{\mu}\lambda_{\mu}^{-1}\frac{\Php_{\mu}(y)}{2}-
\frac{\as^2\,N_c}{\pi}\Psi_{\mu}^{\dagger}(y)\lambda_{\nu}^{-1}\Ra\,\\
&\,&
\Le\Ph_{\nu}(y)-\frac{2\as^2\,N_c}{\pi}
\omega_{\mu}^{-1}\Psi_{\nu}(y)\Ra\,.
\eeqar
For the further simplification of the expression \eq{FP14}
we will do the following. First of all let us consider
the  $\Psi_{\nu}\,,\Psi_{\nu}^{\dagger}$ functions in the \eq{FP14}.
From the \eq{FP6} we have :
\beq\label{FP15}
\sum_{\mu,\nu}\int\,
\frac{d^{2}r_{2}\,r_{13}^{2}}{r_{12}^{2}\,r_{23}^{2}}\,
E_{\mu}(r_{1},r_{2})\,E_{\nu}(r_{2},r_{3})\,
\Ph_{\mu}(y)\,\Ph_{\nu}(y)\,=\,
\sum_{\mu}\,E_{\mu}(r_{1},r_{3})\,\Psi_{\mu}(y)\,,
\eeq
and using the orthonormalization properties
of $E_{\mu}$, \cite{lip1}, we obtain:
\beq\label{FP16}
\Psi_{\mu}(y) = 
\sum_{w,\nu}\int\,\frac{d^{2}r_{2}\,d^{2}r_{1}\,d^{2}r_{3}}
{r_{12}^{2}\,r_{23}^{2}\,r_{13}^{2}}\,
E_{w}(r_{1},r_{2})\,E_{\nu}(r_{2},r_{3})\,
E_{\mu}^{*}(r_{1},r_{3})\,
\Ph_{w}(y)\,\Ph_{\nu}(y)\,=\,
V_{\tilde{\mu},w,\nu}\,
\Ph_{w}(y)\,\Ph_{\nu}(y)\,,
\eeq
where $V_{\mu,w,\nu}$ is the triple pomeron vertex in the conformal
basis, see \cite{Korch1}, and summation over repeating indexes $w,\,\nu$ is assumed
in the form of \eq{Lan21}.
Another observation is concerning the omitted
integration  $\int\frac{d^{2}r_{1}\,d^{2}r_{3}}{r_{13}^{4}}\,$.
This integration over conformal functions 
$E_{\mu}^{*}(r_{1},r_{3})\,E_{\nu}(r_{1},r_{3})$
in the \eq{FP14} gives $\delta_{\mu\nu}$, that, excepting the integration
over rapidity, determines the full Hamiltonian:
\beq\label{FP17}
H=\sum_{\mu}\omega_{\mu}\lambda_{\mu}^{-1}
\Le\Ph_{\mu}(y)-\frac{2\as^2\,N_c}{\pi}
\omega_{\mu}^{-1} V_{\tilde{\mu},w,\nu}
\Ph_{w}(y)\Ph_{\nu}(y)\Ra
\Le\Php_{\mu}(y)-\frac{2\as^2\,N_c}{\pi}
\omega_{\mu}^{-1} V_{\mu,\tilde{w},\tilde{\nu}}
\Php_{w}(y)\Php_{\nu}(y)\Ra\,,
\eeq
in full analogy with the RFT-0 Hamiltonian \eq{FP2}.
We see now, that the "toy" four pomeron vertex in the conformal basis
has the form
\beq\label{FP18}
V_{4P}\,=\,\omega_{\mu}^{-1}\lambda_{\mu}^{-1}\,
\Le\frac{2\as^2\,N_c}{\pi}\Ra^{2}\,
\sum_{w,\nu,w^{'},\nu^{'}}
\Ph_{w}(y)\Ph_{\nu}(y)\,V_{\tilde{\mu},w,\nu}\,
V_{\mu,\tilde{w}^{'},\tilde{\nu}^{'}}\,
\Php_{w^{'}}(y)\Php_{\nu^{'}}(y)\,
\eeq
and the same vertex for the  action in the usual field basis 
looks as follows:
$$
S_{4P}\,=\,-\,2\,\Le\frac{\as^2\,N_c}{\pi}\Ra^{2}\,\int\,dy\int\,
d^{2}r_{1}\,d^{2}r_{3}\int\,
\frac{d^{2}r_{2}\,d^{2}r_{2}^{'}\,}
{r_{12}^{2}\,r_{23}^{2}\,r_{12^{'}}^{2}\,r_{2^{'}3}^{2}}\,
$$
\[
\left\{
\,\Le\,L_{13}\Ph(y,r_{1},r_{2})\,\Ph(y,r_{2},r_{3})\Ra\,
\Le\,H^{-1}(r_{1},r_{3})\Php(y,r_{1},r_{2}^{'})\,\Php(y,r_{2}^{'},r_{3})\Ra\,+\,
\right.
\] 
\beq\label{FP19}
\left.
\,+\,
\Le\,H^{-1}(r_{1},r_{3})\Ph(y,r_{1},r_{2})\,\Ph(y,r_{2},r_{3})\Ra\,
\Le\,L_{13}\Php(y,r_{1},r_{2}^{'})\,\Php(y,r_{2^{'}},r_{3})\Ra\
\right\}\,.
\eeq
From the form of this vertex it is clear, that at least for this 
"toy" four pomeron vertex in the usual basis, the procedure
described in the previous section gives very complicated
auto correlator for the noise field.
Indeed, now, due to the very complicated form of the kernel in the
\eq{FP19}, this auto correlator will include
a complicated expression with the $L$ and $H^{-1}$ operators
with the square of the $\Ph(y,r_{i},r_{j})$ field.
Therefore,  we will write this expression only in Appendix C.
Nevertheless,  for the action in the conformal basis 
the possible expression for the auto correlator of the 
noise field looks much simpler  and we will consider
this derivation in the next section.

\subsection{The Langevin equation for the "toy" four pomeron vertex in the 
conformal basis} 

The action of the effective pomeron field theory 
in the conformal basis looks as follows:
\[
S\,=\,\int\,\,dy\,\sum_{\mu}\,
\left\{
\frac{1}{2}\,\Php_{\mu}(y)\,\lambda_{\mu}^{-1}\,
\frac{\D \Ph_{\mu}(y)}{\D y}\,-\,\frac{1}{2}\,
\Ph_{\mu}(y)\,\lambda_{\mu}^{-1}\,
\frac{\D \Php_{\mu}(y)}{\D y}\,-\,
\right.
\] 
\beq\label{Cnf1}
\left.
-\,\omega_{\mu}\lambda_{\mu}^{-1}
\Le\Ph_{\mu}(y)-\frac{2\as^2\,N_c}{\pi}
\omega_{\mu}^{-1} V_{\tilde{\mu},w,\nu}
\Ph_{w}(y)\Ph_{\nu}(y)\Ra
\Le\Php_{\mu}(y)-\frac{2\as^2\,N_c}{\pi}
\omega_{\mu}^{-1} V_{\mu,\tilde{w},\tilde{\nu}}
\Php_{w}(y)\Php_{\nu}(y)\Ra\right\}\,,
\eeq
where we used the kinematic part of the action from the equation
\eq{FP3}. The action describes the interaction of infinite number
of the pomerons through infinite number of different 
triple pomeron and four pomeron vertexes
in accordance with the results of \cite{braun3,braun4}.
Let us now write the part of the action which contains
the square of the $\Php$ field:
\[
S_{\Ph}\,=\,\int\,dy\,\sum_{\mu}\Le
\frac{2\as^2\,N_c}{\pi}\,
\lambda_{\mu}^{-1}\,
\Ph_{\mu}(y)\,V_{\mu,\tilde{w},\tilde{\nu}}\Php_{w}(y)\Php_{\nu}(y)\,-\,
\right.
\] 
\beq\label{Cnf2}
\left.
\,-\,\Le\frac{2\as^2\,N_c}{\pi}\Ra^{2}\,
\lambda_{\mu}^{-1}\,\omega_{\mu}^{-1}\,
\Ph_{w^{'}}(y)\Ph_{\nu^{'}}(y)\,V_{\tilde{\mu},w^{'},\nu^{'}}\,
V_{\mu,\tilde{w},\tilde{\nu}}\Php_{w}(y)\Php_{\nu}(y)\Ra
\eeq
Proceeding as before we obtain for the $S_{\Ph}$:
\[
S_{\Ph}\,=\,\frac{1}{2}\,
\sum_{\mu}\,
\int\,dy\,\Php_{w}(y)
\Le
\frac{4\as^2\,N_c}{\pi}\,
\lambda_{\mu}^{-1}\,
\Ph_{\mu}(y)\,V_{\mu,\tilde{w},\tilde{\nu}}\,-\,
\right.
\] 
\beq\label{Cnf3}
\left.
\,-\,2\Le\,\frac{2\as^2\,N_c}{\pi}\Ra^{2}\,
\lambda_{\mu}^{-1}\,\omega_{\mu}^{-1}\,
\Ph_{w^{'}}(y)\Ph_{\nu^{'}}(y)\,V_{\tilde{\mu},w^{'},\nu^{'}}\,
V_{\mu,\tilde{w},\tilde{\nu}}\Ra\,\delta(y-y^{'})\,
\int\,dy^{'}\,\Php_{\nu}(y^{'})\,
\eeq
and for the auxiliary field action we have:
\[
S_{Aux}\,=\,-\frac{1}{2}\,
\int\,dy\,\psi_{w}(y)\left\{\sum_{\mu}\,
\Le
\frac{4\as^2\,N_c}{\pi}\,
\lambda_{\mu}^{-1}\,
\Ph_{\mu}(y)\,V_{\mu,\tilde{w},\tilde{\nu}}\,-\,
\right.\right.
\] 
\beq\label{Cnf4}
\left.\left.
\,-\,2\Le\,\frac{2\as^2\,N_c}{\pi}\Ra^{2}\,
\lambda_{\mu}^{-1}\,\omega_{\mu}^{-1}\,
\Ph_{w^{'}}(y)\Ph_{\nu^{'}}(y)\,V_{\tilde{\mu},w^{'},\nu^{'}}\,
V_{\mu,\tilde{w},\tilde{\nu}}\Ra\right\}^{-1}\,\psi_{\nu}(y)\,.
\eeq
Now, whole action \eq{Cnf1} may be written as
\[
S\,=\,\int\,\,dy\,\sum_{\mu}\,
\left\{
\frac{1}{2}\,\Php_{\mu}(y)\,\lambda_{\mu}^{-1}\,
\frac{\D \Ph_{\mu}(y)}{\D y}\,-\,\frac{1}{2}\,
\Ph_{\mu}(y)\,\lambda_{\mu}^{-1}\,
\frac{\D \Php_{\mu}(y)}{\D y}\,-\,
\right.
\] 
\beq\label{Cnf5}
\left.
\,-\,\omega_{\mu}\lambda_{\mu}^{-1}\Ph_{\mu}(y)\Php_{\mu}(y)\,+
\frac{2\as^2\,N_c}{\pi}
\lambda_{\mu}^{-1}\Ph_{w}(y)\Ph_{\nu}(y)\,
V_{\tilde{\mu},w,\nu}\Php_{\mu}(y)\,\right\}+\,S_{Aux}\,-\,
\sum_{\mu}\,\int\,dy\,\psi_{\mu}(y)\,\Php_{\mu}\,.
\eeq
The equation of motion for each $\Ph_{\mu}$ field in this case has the form
of Langevin equation:
\beq\label{Cnf6}
\frac{\D \Ph_{\mu}(y)}{\D y}\,=\,
\omega_{\mu}\,\Ph_{\mu}(y)\,-\,\sum_{w,\nu}\,
\Ph_{w}(y)\Ph_{\nu}(y)\,
V_{\tilde{\mu},w,\nu}\,+\,\psi_{\mu}\,,
\eeq
where we are not summing up over the index $\mu$.
The $\psi_{\mu}$ field in \eq{Cnf6} we can consider as a noise 
field with the following correlators:
\beqar\label{Cnf7}
<\,\psi_{\mu}(y)\,>&=&0\,;\\
<\,\psi_{w}(y),\psi_{\nu}(y_1)\,>&=&
\frac{4\as^2\,N_c}{\pi}\,
\sum_{\mu}\,\lambda_{\mu}^{-1}\,\\
&\,&
\Le\,
\Ph_{\mu}(y)\,V_{\mu,\tilde{w},\tilde{\nu}}
-\frac{2\as^2\,N_c}{\pi}\omega_{\mu}^{-1}
\Ph_{w^{'}}(y)\Ph_{\nu^{'}}(y)\,V_{\tilde{\mu},w^{'},\nu^{'}}
V_{\mu,\tilde{w},\tilde{\nu}}\Ra\,\delta(y-y_1)\,.
\eeqar
We see, that we obtained the infinite number of Langevin equations 
for the fields $\Ph_{\mu}$ 
with  very complicated non diagonal auto correlator for the noise field 
$\psi_{\mu}$. It must be underlined , that the  evolution equations for the 
$\Ph_{\mu}$ fields are also very complicated , there we have
summation over $w,\nu$ indexes in \eq{Cnf6} that means
integration and infinite summation due the  definition of the 
summation procedure in \eq{Lan21}.

\section{The Langevin equation in the dipole approach}

 Now we will consider the results on Langevin equation obtained
in \cite{stoch2} in the framework of the dipole CGC approach
and will reproduce our Langevin equation  
basing on the equations of \cite{stoch2}.
Let us consider the Langevin equation of
\cite{stoch2} for the process of 
the scattering of an arbitrary
numbers of dipoles off the target. 
The noise (fluctuation) term of the Langevin equation, corresponding
to this process,  was obtained in  \cite{stoch2}
and it has the following form: 
\beq\label{Dip1}
f(r_1,r_2,y)\,=\,C\,\int\,d^{2}\rho_{1}\,d^{2}\rho_{2}\,d^{2}\rho_{3}\,
G_{0}(r_1,r_2|\rho_{1},\rho_{3})\,\frac{|\rho_{12}|}{\rho_{13}^{2}}\,
\sqrt{\nabla^{2}_{\rho_{1}}\,\nabla^{2}_{\rho_{2}}
\Ph(y,\rho_{1},\rho_{2})}\,
\nu(\rho_{1},\rho_{2},\rho_{3},y)\,
\eeq
where we adapted the notations of \cite{stoch2}
on the notations of the paper, here
we introduced a $\Ph(r_i,r_j)$ field  as a RFT counterpart 
of the  $T_{Y}(r_i,r_j)$ 
function from \cite{stoch2}. The constant C in the 
expression \eq{Dip1} is related to the
possible difference in normalization of the amplitudes 
in both papers and we do not 
fix it in the expression \eq{Dip1}, for our following 
derivation it is not important.  
The noise field $\nu(\rho_{1},\rho_{2},\rho_{3},y)$ in \eq{Dip1}
is a noise field from \cite{stoch2} with the following auto correlator:
\beq\label{Dip2}
<\nu(\rho_{1},\rho_{2},\rho_{3},y)\,,\,\nu(\rho_{1}^{'},
\rho_{2}^{'},\rho_{3}^{'},y^{'})>\,=\,
\delta^{2}(\rho_{1}-\rho_{2}^{'})\,
\delta^{2}(\rho_{2}-\rho_{1}^{'})\,
\delta^{2}(\rho_{3}-\rho_{3}^{'})\,
\delta(y-y^{'})\,.
\eeq
Using \eq{Dip2},  we can 
define the auto correlator for the field $f(r_i,r_j,y)$:
\beq\label{Dip3}
< f(r_1,r_2,y)\,,\,f(r_{1}^{'},r_{2}^{'},y) >\, = \,
C^{2}\,\int\,d^{2}\rho_{1}\,d^{2}\rho_{2}\,d^{2}\rho_{3}\,
\frac{|\rho_{12}|}{\rho_{13}^{2}}\,
G_{0}(r_1,r_2|\rho_{1},\rho_{3})\,
\sqrt{\nabla^{2}_{\rho_{1}}\,\nabla^{2}_{\rho_{2}}\,
\Ph(y,\rho_{1},\rho_{2})}\,
\eeq
$$
\int\,d^{2}\rho_{1}^{'}\,d^{2}\rho_{2}^{'}\,d^{2}\rho_{3}^{'}\,
\frac{|\rho_{1^{'}2^{'}}|}{\rho_{1^{'}3^{'}}^{2}}\,
G_{0}(r_{1}^{'},r_{2}^{'}|\rho_{1}^{'},\rho_{3}^{'})\,
\sqrt{\nabla^{2}_{\rho_{1}^{'}}\,\nabla^{2}_{\rho_{2}^{'}}
\Ph(y,\rho_{1}^{'},\rho_{2}^{'})}
<\nu(\rho_{1},\rho_{2},\rho_{3},y)\,,\,\nu(\rho_{1}^{'},
\rho_{2}^{'},\rho_{3}^{'},y^{'})>\,\,.
$$
The straightforward calculations gives with the help of \eq{Dip2}:
\beq\label{Dip4}
< f(r_1,r_2,y)\,,\,f(r_{1}^{'},r_{2}^{'},y^{'}) >\,=\,
C^{2}\,\int\,d^{2}\rho_{1}\,d^{2}\rho_{2}\,d^{2}\rho_{3}\,
G_{0}(r_1,r_2|\rho_{1},\rho_{3})\,
G_{0}(r_{1}^{'},r_{2}^{'}|\rho_{2},\rho_{3})\,
\eeq
$$
\sqrt{\nabla^{2}_{\rho_{1}}\,\nabla^{2}_{\rho_{2}}\,
\Ph(y,\rho_{1},\rho_{2})}\,
\sqrt{\nabla^{2}_{\rho_{2}}\,\nabla^{2}_{\rho_{1}}\,
\Ph(y,\rho_{2},\rho_{1})}\,
\frac{\rho_{12}^{2}}{\rho_{13}^{2}\,\rho_{23}^{2}}\,
\delta(y-y^{'})\,.
$$
Now we will use \eq{Lan13} in order to rewrite Green's functions
$G_{0}$ in the following form:
\beq\label{Dip5}
G_{0}(r_1,r_2|\rho_{1},\rho_{3})\,=\,
\nabla^{-2}_{r_{1}}\,\nabla^{-2}_{r_{2}}\,
\delta^{2}(r_{1}-\rho_{1})\,\delta^{2}(r_{2}-\rho_{3})\,
\eeq
and 
\beq\label{Dip6}
G_{0}(r_{1}^{'},r_{2}^{'}|\rho_{2},\rho_{3})\,=\,
\nabla^{-2}_{r_{1}^{'}}\,\nabla^{-2}_{r_{2}^{'}}\,
\delta^{2}(r_{1}^{'}-\rho_{2})\,\delta^{2}(r_{2}^{'}-\rho_{3})\,.
\eeq
Inserting \eq{Dip5} and  \eq{Dip6} into the \eq{Dip4}
it is easy to see, that operators 
$\nabla^{-2}_{r_{1}}\,\nabla^{-2}_{r_{2}}\,$ and 
$\nabla^{-2}_{r_{1}^{'}}\,\nabla^{-2}_{r_{2}^{'}}\,$
now may be reabsorbed in the definition of our initial field
$f(r_i,r_j,y)$:
\beq\label{Dip7}
f(r_i,r_j,y)\,\rightarrow\,
\nabla^{-2}_{r_{i}}\,\nabla^{-2}_{r_{j}}\,\tilde{f}(r_i,r_j,y)\,.
\eeq
With this new noise field $\tilde{f}(r_i,r_j,y)$, the equation of motion
for the $\Ph(r_i,r_j)$ field, or for the $T_{Y}(r_i,r_j)$ 
amplitude from \cite{stoch2}, obtains precisely
the same additional noise term as in \eq{Lan23}. Indeed, now the auto 
correlator for the  $\tilde{f}(r_i,r_j,y)$ has the form:
\beq\label{Dip8}
< \tilde{f}(r_1,r_2,y)\,,\,
\tilde{f}(r_{1}^{'},r_{2}^{'},y^{'}) >\,=\,
C^{2}\,\int\,d^{2}\rho_{1}\,d^{2}\rho_{2}\,d^{2}\rho_{3}\,
\sqrt{\nabla^{2}_{\rho_{1}}\,\nabla^{2}_{\rho_{2}}\,
\Ph(y,\rho_{1},\rho_{2})}\,
\sqrt{\nabla^{2}_{\rho_{2}}\,\nabla^{2}_{\rho_{1}}\,
\Ph(y,\rho_{2},\rho_{1})}\,
\eeq
$$
\frac{\rho_{12}^{2}}{\rho_{13}^{2}\,\rho_{23}^{2}}\,
\delta^{2}(r_{1}-\rho_{1})\,\delta^{2}(r_{2}-\rho_{3})\,
\delta^{2}(r_{1}^{'}-\rho_{2})\,\delta^{2}(r_{2}^{'}-\rho_{3})\,
\delta(y-y^{'})\,.
$$
Performing integration and interchanging variables 
$r_{1}^{'}$ and $r_{2}^{'}$ in the 
$\tilde{f}(r_{1}^{'},r_{2}^{'},y^{'})$, we obtain:
\beq\label{Dip9}
< \tilde{f}(r_1,r_2,y)\,,\,
\tilde{f}(r_{2}^{'},r_{1}^{'},y^{'}) >\,=\,
C^{2}\,\frac{r_{11^{'}}^{2}}{r_{12}^{2}\,r_{1^{'}2^{'}}^{2}}\,
\Le\,
\nabla^{2}_{r_{1}}\,\nabla^{2}_{r_{1^{'}}}\,
\Ph(y,r_{1},r_{1}^{'})\,\Ra\,\delta^{2}(r_{2}-r_{2}^{'})\,
\delta(y-y^{'})\,.
\eeq
Comparing this expression with the  auto correlator given by \eq{Lan9}:
\beq\label{Dip10}
<\,\psi(y,\rho_{1},\rho_{2})\,,
\psi(y^{'},\rho_{2}^{'},\rho_{1}^{'})\,>\,= \,
\frac{4\as^2\,N_c}{\pi}\,
\frac{\rho^{2}_{12}\,\rho^{2}_{1^{'}2^{'}}}{\rho^{2}_{11^{'}}}\,
(L_{11^{'}}\,\Ph(y,\rho_{1},\rho_{1}^{'}))\,
\delta(y-y^{'})\,
\delta^{2}(\rho_{2}-\rho_{2}^{'})\,,
\eeq
we see, that if
we reabsorb $1/\rho_{ij}^{4}$ coefficient in definition
of the noise field $\psi(y,\rho_{i},\rho_{j})$ in \eq{Dip10}, then
we obtain the same as 
\eq{Dip9} auto correlator for the noise field.

\section{Discussion of results}

 As a first main result of the paper
we consider a  
Langevin equations 
\eq{Lan8}-\eq{Lan9}, \eq{Phi11}-\eq{Phi12} and \eq{Cnf6}-\eq{Cnf7}
obtained in the framework of QCD-RFT.
The meaning of these equations is simple.
The derivation of them is achieved with the use of
the classical action of the theory,
i.e.  action without pomeron loops . The introduction of the 
auxiliary field in the classical action instead the 
square of the $\Php$ field
leads to the Langevin equation for the field $\Ph$.
Iterating calculations procedure for the  
$\Ph$ field together with the calculation of the 
auto correlator for the noise field
$\psi$, see for example discussion in \cite{stoch1}, 
will lead to the account of the pomeron loops contribution 
in the pomeron field $\Ph$.
Proceeding, we will calculate an
all loops contribution to the  pomeron field and will resolve whole 
quantum problem for this effective theory. Of course,
practically, due the
very complicated form of the auto correlators for the noise fields, this task looks 
at least not easy.

 The interpretation of the Langevin equations in this paper and
the interpretation of the Langevin equations from 
the papers \cite{stoch1,stoch2} are the same,
in spite to the different forms of obtained equations. 
As in \cite{stoch1} and \cite{stoch2} 
we also obtained two different Langevin equations formulated
in the different  frameworks.
Let us , therefore,  discuss the similarities and 
differences in the forms of the Langevin equations,
formulated in the momentum and coordinate spaces
separately.
First of all, we consider the obtained Langevin equation 
\eq{Phi11}
for the $\varphi(k,y)$ field in the momentum space.
The \eq{Phi11} looks very similar
to the Langevin equations of \cite{stoch1}. There is only one move which
reduces \eq{Phi11} to the corresponding equations in \cite{stoch1} and
this move is an assumption about the action of the operators
$L$ and $L^{-1}$ of \eq{EFT4}, 
on the function $\varphi(k,y)$. If we assume, that
the $\varphi(k,y)$ is a eigenfunction of $L$ and $L^{-1}$ with 
the unit eigenvalues then the corresponding equations will be the same.
But, as it was argued in this paper, this move is impossible
in the framework of the present approach because in this case
a very unphysical
restriction \eq{Phi13} on the form of the unintegrated gluon density function
$f(y,k)$ is arising. Therefore, in general, without some
approximation it is impossible to write \eq{Phi11} in the form obtained in
\cite{stoch1}. More important, that 
the function $\varphi(k,y)$ is defined for the case of zero momentum transfer,
i.e. there is no correct impact parameter dependence account of the amplitude
is performed, 
and, therefore, it is in principal impossible describe 
loops contribution in the pomeron field with
the use of the  $\varphi(k,y)$ field. 
The Langevin equation for this function in the form
of \eq{Phi11}, therefore,  may plays a role of some toy model which probably
may describe some properties of real QCD. 
In general, this function
also may be applied for the semiclassical solution of the problem, i.e.
for the calculation of the ''tree'' pomeron structure, see \cite{bond1}. 
Another application of this amplitude and semiclassical  approach 
was developed in \cite{LevLu1,lev1}, where the calculations 
of the amplitude were performed in the
framework of a generating functional approach which has a strong relations with the 
probabilistic interpretation of the BFKL pomeron interactions.
It must be also noticed, that the approach developed in \cite{lev1}
has a mutual roots with the approach considered in this paper, 
focusing, nevertheless, mostly on the semiclassical solutions of the problem.

  Concerning the correct Langevin equation for the $\Ph(y,r_{i},r_{j})$
field in the transverse position space
we see, that the situation here is more obvious. The calculations in  the 
previous section show the equivalence between the Langevin equation
obtained in \cite{stoch2} in the framework of the s-channel dipole model
and Langevin equation given by \eq{Lan8}-\eq{Lan9} which was obtained
in the framework of the t-channel QCD-RFT. In spite to the 
differently written
forms of the noise terms and auto correlators of
the noise terms , the Langevin equations in both approaches are the same.
The Langevin equations it is a another description
of the dynamics of the physical problem, and the same form of the
Langevin equations in both approaches
means a equivalent description of the quantum fields  dynamics
in QCD-RFT and dipole approaches. Using Langevin equation as a 
bridge we can pass from the QCD-RFT side to the dipole side
of extended Balitsky-JIMWLK hierarchy and vice versa.
We can conclude, therefore, that the $T_{Y}(r_i,r_j)$ 
function considered in \cite{stoch2} in the framework of the  
extended Balitsky-JIMWLK hierarchy of equations
is equivalent to the pomeron field considered in the framework of
t-channel QCD-RFT, where in both cases all pomeron loops 
contribution is included in. More of that,
due the equivalence between the QCD-RFT approach and dipole 
model of \cite{stoch2} prooven in this paper and due the fact that
in \cite{stoch2} was established the equivalence between
approaches of \cite{stoch2} and \cite{stoch3}, we can conclude, that
our QCD RFT t-channel model is equivalent to the modified version
of the JIMWLK equation of \cite{stoch3} as well.
On the level of RFT in zero dimensions
this fact was clarified in \cite{bond2}, and now we see,
that the s-channel dipole  and t-channel RFT  approaches are indeed  equivalent 
in the physical space
of two transverse dimensions. This result we consider as a second main result
of the paper.

 The consideration of the theory with
the "toy" four pomeron vertex in the conformal basis
is similar, as it must be, to the derivations of \cite{braun3,braun4}.
As in \cite{braun3}, we obtained the reformulated theory
with infinite number of one dimensional pomerons,
with only additional terms due the four pomeron vertex. 
In spite to the redefinition of the degrees
of the freedom of the theory, the corresponding equations are
not simpler then the equations in the usual basis.
Indeed, the equation of motion of this theory
contains infinite number of one dimension pomerons with an 
infinite number of different and very
complicated interaction vertexes which include three
integration and three infinite summation over repeating indexes.
The equations in the form of Langevin dynamics
also include very complicated correlators
of \eq{Cnf7}, where the auto correlator for given
conformal weights $\mu\,,\nu$ depends on  
all other conformal fields $\Ph$ of the theory. 
Nevertheless, as it was mentioned in \cite{braun3},
we can truncate the sums on some value of conformal weight $\mu$
and try to solve simplified truncated theory. Still,
this task will be not easy and we leave it for the future studies.
Concerning the form of the 
expression for this four pomeron vertex 
it is also interesting to note, that obtained in this paper
vertex is the same as the 
four point function of two-dimensional
conformal field theory derived in \cite{Korch1}
from the conformal bootstrap point of view on the
BFKL pomeron. In the given context it means, that
conformal bootstrap in application to the interacting  
BFKL pomerons will lead to the factorized Hamiltonian 
of the problem with precisely zero ground state,
see beginning of the Section 4. We do not know about
the deep physical reasons for a such relation
between  the conformal field theory in two dimensions and the
Hamiltonian of the interacting BFKL pomerons, and, therefore,
we do not consider here this subject, which potentially
may be very interesting. Finishing the theme of four 
pomeron interaction vertex, we can also notice,
that our vertex is not the same as the four pomeron vertex
introduced in the \cite{LevLu1} on the base of the physical reasons
and arguments different from the considered in this paper.

\section*{Acknowledgments}
I am  especially grateful to Leszek Motyka and Mikhail Braun  
for the  discussion
on the subject of the paper. I thank also 
E.Levin, N.Armesto and C.Pajares for the interesting discussions and
useful comments 
I also gratefully acknowledge 
the support of the Ministerio de Educacion y Ciencia of Spain
under project FPA2005-01963, and by Xunta de Galicia 
(Conselleria de Educacion).

\vspace{3mm}

\newpage
\section*{Appendix A:}

\renewcommand{\theequation}{A.\arabic{equation}}
\setcounter{equation}{0}

 Let us consider the following part of the action:
\beq\label{Ap1}
\hat{S}_I \, =\,
\int\,dy\,\int\,d^{2}r_{1}\,d^{2}r_{2}\,d^{2}r_{3}\,
\,(\frac{L_{13}}{\raddd}\,(\raddd  \php(y,r_{1},r_{3})))\,
\phi(y,r_{1},r_{2})\,\phi(y,r_{2},r_{3})\,,
\eeq
where we omitted the coefficient in front of \eq{EFT17}.
Let us make the Fourier transform of the $\phi$ functions:
\beq\label{Ap2}
\phi(y,r_{1},r_{2})\,=\,
\int\,\frac{d^2\,k_1\,d^2\,q_1\,}{(2\,\pi)^2}\,
e^{-i\,r_{1}\,k_1\,-i\,r_{2}\,(q_1\,-\,k_1)}\,
\tilde{\phi}(y,k_1,q_1-k_1)\,,
\eeq
$$
\phi(y,r_{2},r_{3})\,=\,
\int\,\frac{d^2\,k_2\,d^2\,q_2\,}{(2\,\pi)^2}\,
e^{-i\,r_{2}\,k_2\,-i\,r_{3}\,(q_2\,-\,k_2)}\,
\tilde{\phi}(y,k_2,q_2-k_2)\,,
$$
$$
\php(y,r_{1},r_{3})\,=\,
\int\,\frac{d^2\,k_3\,d^2\,q_3\,}{(2\,\pi)^2}\,
e^{i\,r_{3}\,k_3\,+i\,r_{1}\,(q_3\,-\,k_3)}\,
\tilde{\php}(y,k_3,q_3-k_3)\,.
$$
Operator $\frac{L_{13}}{\raddd}\,\raddd$ throw the Fourier transform
is changed to :
\beq\label{Ap3} 
\frac{L_{13}}{\raddd}\,\raddd\,=\,r^{2}_{13}\,
\nabla^{2}_{1}\,\nabla^{2}_{3}\,r^{2}_{13}\,\stackrel{FT}{\rightarrow}\,
\,\nabla^{2}_{k_3}\,k^{2}_{3}\,
(q_{3}-k_{3})^{2}\,\nabla^{2}_{k_3}\,=\,\hat{L}_{3}\,.
\eeq
Inserting \eq{Ap2} and \eq{Ap3} back in \eq{Ap1} we obtain:
\beq\label{Ap4}
\hat{S}_I =
\int\,dy\int\,d^{2}k_{1}\,d^{2}k_{2}\,d^{2}k_{3}
\int\,d^{2}q_{1}\,d^{2}q_{2}\,d^{2}q_{3}
(\hat{L}_{3}\,\tilde{\php}(y,k_{3},q_{3}-k_{3}))
\tilde{\phi}(y,k_{2},q_{2}-k_{2})
\tilde{\phi}(y,k_{1},q_{1}-k_{1})
\eeq
$$
\int\,\frac{d^{2}r_{1}}{(2\pi)^2}\,
\frac{d^{2}r_{2}}{(2\pi)^2}\,
\frac{d^{2}r_{3}}{(2\pi)^2}\,
e^{-i\,r_{1}\,k_1\,-i\,r_{2}\,(q_1\,-\,k_1)}\,
e^{-i\,r_{2}\,k_2\,-i\,r_{3}\,(q_2\,-\,k_2)}\,
e^{i\,r_{3}\,k_3\,+i\,r_{1}\,(q_3\,-\,k_3)}\,.
$$
Let us consider the last line of \eq{Ap4}:
\beq\label{Ap5}
\int\,\frac{d^{2}r_{1}}{(2\pi)^2}\,
\frac{d^{2}r_{2}}{(2\pi)^2}\,
\frac{d^{2}r_{3}}{(2\pi)^2}\,
e^{-i\,r_{1}\,k_1\,-i\,r_{2}\,(q_1\,-\,k_1)}\,
e^{-i\,r_{2}\,k_2\,-i\,r_{3}\,(q_2\,-\,k_2)}\,
e^{i\,r_{3}\,k_3\,+i\,r_{1}\,(q_3\,-\,k_3)}\,
\eeq
and make there the following change of variables:
\beqar\label{Ap6}
\rho_{1}=\frac{r_{3}-r_{1}}{2}\,\,& &
r_{3}=\rho_{1}+\rho_{2}\,\\
\rho_{2}=\frac{r_{3}+r_{1}}{2}\,\,& &
r_{1}=\rho_{2}-\rho_{1}\,\\
R=\frac{r_{2}}{2}-\frac{r_{1}}{4}-\frac{r_{3}}{4}\,\,& &
r_{2}=2R+\rho_{2}\,.
\eeqar
Changing variables in \eq{Ap5} we obtain:
\beq\label{Ap7}
4\int\,\frac{d^{2}R}{(2\pi)^2}\,
\frac{d^{2}\rho_{1}}{(2\pi)^2}\,
\frac{d^{2}\rho_{2}}{(2\pi)^2}\,
e^{i\,R(k_1\,-\,q_1\,-\,k_2)}\,
e^{i\,\rho_{1}(\,k_1\,+\,k_2\,+\,2\,k_3-\,q_2\,-\,q_3)}\,
e^{i\,\rho_{2}(q_3\,-\,q_1-\,q_2)}\,.
\eeq
Integration over $R$,$\rho_{1}$ and $\rho_{2}$
gives three delta function and after the integration over
$q_1$, $k_1$ and $q_2$ such that
\beqar\label{Ap8}
q_1\,& = &\,q_3\,-\,k_2\,-\,k_3\,\\
k_1\,& = &\,q_3\,-\,k_3\,\\
q_2\,& = &\,k_2\,+\,k_3\,
\eeqar
we obtain \eq{EFT9}.

\section*{Appendix B:}

 \renewcommand{\theequation}{B.\arabic{equation}}
\setcounter{equation}{0}

 Let us consider together \eq{EFT13} and \eq{EFT18}:
\beq\label{Ap9}
\hat{S}_I \,=\,
\int\,d^{2}k\,
\,(\,\nabla^{2}_{k}\,k^{4}\,
\,\nabla^{2}_{k}\,\varphi^{\dagger}(y,k))\,
\varphi(y,k)\,\varphi(y,k)\,,
\eeq
and
\beqar\label{Ap10}
\varphi(y,k)\,& = &\,{\pi^2 \over 2\,N_c\,}\,
\int_{k^2}^{\infty}\,dk^{'2}\,\frac{f(y,k^{'})}{k^{'4}}\,
\log(\frac{k^{'2}}{k^2})\,\\
\varphi^{\dagger}(y,k)\,& = &\,{\pi^2 \over 2\,N_c\,}\,
\int_{k^2}^{\infty}\,dk^{'2}\,\frac{f^{\dagger}(y,k^{'})}{k^{'4}}\,
\log(\frac{k^{'2}}{k^2})\,.
\eeqar
The operator $\nabla^{2}_{k}\,$,
because the rotational invariance of the problem (zero transfer momentum),
can be rewritten in the following form:
\beq\label{Ap11}
\nabla^{2}_{k}\,=\,4\,\frac{\D}{\D\,k^2}(k^2\,\frac{\D}{\D\,k^2})\,=\,
\,4\,\frac{\D}{\D\,x}(x\,\frac{\D}{\D\,x})\,,
\eeq
here we changed the variables: $k^2\,\rightarrow\,x$.
The whole expression \eq{Ap9} now may be rewritten
in the terms of variable x:
\beq\label{Ap12}
\hat{S}_I \,=\,4\,\pi
\int\,dx\,\frac{\D}{\D\,x}(x\,\frac{\D}{\D\,x})
\,\,(x^{2}\,
\,\nabla^{2}_{k}\,\varphi^{\dagger}(x))\,
\varphi(x)\,\varphi(x)\,.
\eeq
Now we  use , that 
\beqar\label{Ap13}
f^{\dagger}(x)\,& \,\propto\,&\,x^{2}\,
\,\nabla^{2}_{k}\,\varphi^{\dagger}(x)\,\\
\varphi(x)\,& \,\propto\, &
\int_{x}^{\infty}\,dx^{'}\,\frac{f(x^{'})}{x^{'2}}\,
\log(\frac{x^{'}}{x})\,.
\eeqar
Inserting these expressions in the \eq{Ap12} we obtain:
\beq\label{Ap14}
\hat{S}_I \,\propto\,
\int\,dx\,\frac{\D}{\D\,x}(x\,\frac{\D}{\D\,x})
\,\,(f^{\dagger}(x))\,
\int_{x}^{\infty}\,dx^{'}\,\frac{f(x^{'})}{x^{'2}}\,
\log(\frac{x^{'}}{x})\,
\int_{x}^{\infty}\,dx^{"}\,\frac{f(x^{"})}{x^{"2}}\,
\log(\frac{x^{"}}{x})\,.
\eeq
Integrating \eq{Ap14} by parts and 
omitting boundary terms of integration we obtain:
\beq\label{Ap15}
\hat{S}_I \,\propto\,2\,
\int\,dx\,(\frac{\D}{\D\,x}
\,f^{\dagger}(x))\,\Le\,x\,
\int_{x}^{\infty}\,dx^{"}\,\frac{f(x^{"})}{x^{"2}}\,
\log(\frac{x^{"}}{x})\,
\int_{x}^{\infty}\,dx^{'}\,\frac{f(x^{'})}{x\,x^{'2}}\,\Ra\,.
\eeq
The result of the second integration by parts is the following:
\beq\label{Ap16}
\hat{S}_I \,\propto\,-2\,
\int\,dx\,f^{\dagger}(x)\,\Le\,
\int_{x}^{\infty}\,dx^{"}\,\frac{f(x^{"})}{x^{"2}}\,
\log(\frac{x^{"}}{x})\,
\int_{x}^{\infty}\,dx^{'}\,\frac{f(x^{'})}{x\,x^{'2}}\,\Ra\,+\,
\eeq
$$
\,+\,2\,
\int\,dx\,f^{\dagger}(x)\,\Le\,x\,
\int_{x}^{\infty}\,dx^{"}\,\frac{f(x^{"})}{x\,x^{"2}}\,
\int_{x}^{\infty}\,dx^{'}\,\frac{f(x^{'})}{x\,x^{'2}}\,\Ra\,+\,
$$
$$
\,+\,2\,
\int\,dx\,f^{\dagger}(x)\,\Le\,x\,
\int_{x}^{\infty}\,dx^{"}\,\frac{f(x^{"})}{x^{"2}}\,
\log(\frac{x^{"}}{x})\,
\Le\,\frac{f(x)}{x^3}+
\int_{x}^{\infty}\,dx^{'}\,\frac{f(x^{'})}{x^2\,x^{'2}}\,\Ra\,\Ra\,,
$$
that gives finally:
\beq\label{Ap17}
\hat{S}_I \,\propto\,2\,
\int\,\frac{dx}{x}\,f^{\dagger}(x)\,
\int_{x}^{\infty}\,dx^{"}\,\frac{f(x^{"})}{x^{"2}}\,
\int_{x}^{\infty}\,dx^{'}\,\frac{f(x^{'})}{x^{'2}}\,+\,
\eeq
$$
\,+\,2\,
\int\,\frac{dx}{x^2}\,f^{\dagger}(x)\,f(x)\,
\int_{x}^{\infty}\,dx^{"}\,\frac{f(x^{"})}{x^{"2}}\,
\log(\frac{x^{"}}{x})\,.
$$
We see that this vertex is the same as in \cite{vert1}, see
also \cite{bond1},
excepting some not important for our consideration 
constant in front of the expression.

\section*{Appendix C:}

 \renewcommand{\theequation}{C.\arabic{equation}}
\setcounter{equation}{0}

 Let us consider expression \eq{FP19}
$$
S_{4P}\,=\,-\,2\,\Le\frac{\as^2\,N_c}{\pi}\Ra^{2}\,\int\,dy\int\,
d^{2}r_{1}\,d^{2}r_{3}\int\,
\frac{d^{2}r_{2}\,d^{2}r_{2}^{'}\,}
{r_{12}^{2}\,r_{23}^{2}\,r_{12^{'}}^{2}\,r_{2^{'}3}^{2}}\,
$$
\[
\left\{
\,\Le\,L_{13}\Ph(y,r_{1},r_{2})\,\Ph(y,r_{2},r_{3})\Ra\,
\Le\,H^{-1}(r_{1},r_{3})\Php(y,r_{1},r_{2}^{'})\,\Php(y,r_{2^{'}},r_{3})\Ra\,+\,
\right.
\] 
\beq\label{Ap18}
\left.
\,+\,
\Le\,H^{-1}(r_{1},r_{3})\Ph(y,r_{1},r_{2})\,\Ph(y,r_{2},r_{3})\Ra\,
\Le\,L_{13}\Php(y,r_{1},r_{2}^{'})\,\Php(y,r_{2}^{'},r_{3})\Ra\
\right\}\,.
\eeq
and let us rewrite this expression in the following form:
$$
S_{4P}\,=\,-\,\frac{1}{2}\,
\int\,dy\,dy^{'}\int\,
\frac{d^{2}r_{1}\,d^{2}r_{3}}{r_{12^{'}}^{4}}
\int\,
\frac{d^{2}r_{2}^{'}\,d^{2}r_{2}^{"}}{r_{2^{"}3}^{4}}\,
\Php(y,r_{1},r_{2}^{'})\,\Php(y,r_{2}^{"},r_{3})\cdot
$$
\[
\cdot\Le\frac{2\as^2\,N_c}{\pi}\Ra^{2}\,\int
\frac{r_{12^{'}}^{2}\,r_{2^{"}3}^{2}}
{r_{12}^{2}\,r_{23}^{2}}\,\,d^{2}r_{2}\,
\left\{\Le L_{13}\Ph(y,r_{1},r_{2})\Ph(y,r_{2},r_{3})\Ra
H^{-1}(r_{1},r_{3})\,+\,
\right.
\] 
\beq\label{Ap19}
\left.
\,+\,
\Le\,H^{-1}(r_{1},r_{3})\Ph(y,r_{1},r_{2})\Ph(y,r_{2},r_{3})\Ra
L_{13}\right\}\delta(y-y^{'})\,\delta^{2}(r_{2}^{'}-r_{2}^{"})
\eeq
From this expression for the $S_{4P}$ part of the action it is clear, that
auto correlator \eq{Lan9} will be changed now to the following form:
$$
<\,\psi(y,r_{1},r_{2}^{'})\,,
\psi(y^{'},r_{2}^{"},r_{3})\,>\, = 
\frac{4\as^2\,N_c}{\pi}\,
\frac{r^{2}_{12^{'}}\,r^{2}_{2^{"}3}}{r^{2}_{12^{"}}}\,
(L_{12^{"}}\,\Ph(y,r_{1},r_{2}^{"}))\,
\delta(y-y^{'})\,
\delta^{2}(r_{2}^{'}-r_{2}^{"})\,-\,
$$
\[
\,-\,
\Le\frac{2\as^2\,N_c}{\pi}\Ra^{2}\,\int
\frac{r_{12^{'}}^{2}\,r_{2^{"}3}^{2}}
{r_{12}^{2}\,r_{23}^{2}}\,\,d^{2}r_{2}\,
\left\{\Le L_{13}\Ph(y,r_{1},r_{2})\Ph(y,r_{2},r_{3})\Ra
H^{-1}(r_{1},r_{3})\,+\,
\right.
\] 
\beq\label{Ap20}
\left.
\,+\,
\Le\,H^{-1}(r_{1},r_{3})\Ph(y,r_{1},r_{2})\Ph(y,r_{2},r_{3})\Ra
L_{13}\right\}\delta(y-y^{'})\,\delta^{2}(r_{2}^{'}-r_{2}^{"})\,,
\eeq
whereas the form of equation of motion \eq{Lan8} will stay unchanged.

\end{document}